\documentclass[reprint,superscriptaddress,
showpacs,preprintnumbers,
notitlepage,amsmath,amssymb,
aps,
prb
]{revtex4-1}

\usepackage{graphicx}
\usepackage{dcolumn}
\usepackage{bm}
\usepackage{tikz}
\usepackage{bbm}
\usepackage{diagbox}
\usepackage{braket}
\usepackage{appendix}
\usepackage{graphicx}
\usepackage{enumerate}
\usepackage{subfigure}
\usepackage{epstopdf}
\usepackage[colorlinks,
linkcolor=red,
anchorcolor=black,
citecolor=blue]{hyperref}
\usepackage{amssymb}
\usepackage{framed}
\usepackage{tabularx}
\usepackage{booktabs}
\usepackage{float}
\usepackage[table]{xcolor}
\usepackage{array}
\usepackage{caption}
\usepackage{tikz}
\usetikzlibrary{positioning}
\usetikzlibrary{decorations,arrows}
\usetikzlibrary{decorations.markings}
\tikzset{>=stealth}
\hypersetup{colorlinks=true}

\newcommand{\1}{\text{\uppercase\expandafter{\romannumeral1}}}
\newcommand{\2}{\text{\uppercase\expandafter{\romannumeral2}}}
\newcommand{\3}{\text{\uppercase\expandafter{\romannumeral3}}}
\newcommand{\4}{\text{\uppercase\expandafter{\romannumeral4}}}
\newcommand{\5}{\text{\uppercase\expandafter{\romannumeral5}}}
\newcommand{\6}{\text{\uppercase\expandafter{\romannumeral6}}}

\newcommand{\trishape}[3]{
\begin{tikzpicture}[scale=#1, baseline=0]
    \draw (0.75*#3,0+#2) -- (-0.75*#3,-0.866+#2) -- (-0.75*#3,0.866+#2) -- cycle;
\end{tikzpicture}%
}

\newcommand{\triright}{\mathchoice
  {\trishape{0.17}{0.55}{1}}
  {\trishape{0.17}{0.55}{1}}
  {\trishape{0.09}{0.5}{1}}
  {\trishape{0.06}{0.07}{1}}
}

\newcommand{\trileft}{\mathchoice
  {\trishape{0.17}{0.55}{-1}}
  {\trishape{0.17}{0.55}{-1}}
  {\trishape{0.09}{0.5}{-1}}
  {\trishape{0.06}{0.07}{-1}}
}

\newcommand{\hexthreelinesaux}[3]{%
\begin{tikzpicture}[scale=#1, baseline=0]
    \begin{scope}[shift={(0,#2)}]
        \foreach \a in {#3} {
            \draw (0,0) -- (\a:1);
        }
    \end{scope}
\end{tikzpicture}%
}

\newcommand{\trilinesU}{\mathchoice
  {\hexthreelinesaux{0.15}{0.3}{90,210,330}}
  {\hexthreelinesaux{0.15}{0.3}{90,210,330}}
  {\hexthreelinesaux{0.11}{0.3}{90,210,330}}
  {\hexthreelinesaux{0.06}{0.07}{90,210,330}}
}

\newcommand{\trilinesD}{\mathchoice
  {\hexthreelinesaux{0.15}{0.65}{270,150,30}}
  {\hexthreelinesaux{0.15}{0.65}{270,150,30}}
  {\hexthreelinesaux{0.11}{0.7}{270,150,30}}
  {\hexthreelinesaux{0.06}{0.07}{270,150,30}}
}

\newcommand{\hexlines}{\mathchoice
  {\hexthreelinesaux{0.15}{0.65}{30,90,150,210,270,330}}
  {\hexthreelinesaux{0.15}{0.65}{30,90,150,210,270,330}}
  {\hexthreelinesaux{0.11}{0.7}{30,90,150,210,270,330}}
  {\hexthreelinesaux{0.06}{0.07}{30,90,150,210,270,330}}
}

\newcommand{\hexcenterdot}{\mathchoice
  {\hexcenterdotaux{0.35}{-0.8ex}}   
  {\hexcenterdotaux{0.35}{-0.8ex}}   
  {\hexcenterdotaux{0.12}{-0.6ex}}   
  {\hexcenterdotaux{0.09}{-0.5ex}}   
}

\newcommand{\hexcenterdotaux}[2]{%
\begin{tikzpicture}[scale=#1, baseline=#2, rotate=0]
    \draw (90:1) \foreach \x in {1,...,5} { -- (90+60*\x:1) } -- cycle;
    \fill (0,0) circle (0.25);
\end{tikzpicture}%
}

\begin{document}
\title{Engineering Tunable Kagome Moir\'e Superlattices in Twisted Transition Metal Dichalcogenides}
\author{Adrian Fedorko}
\affiliation{Department of Physics, The Pennsylvania State University, University Park, Pennsylvania 16802, USA}
\author{Chao-Xing Liu}
\affiliation{Department of Physics, The Pennsylvania State University, University Park, Pennsylvania 16802, USA}
\affiliation{Center for Theory of Emergent Quantum Matter, The Pennsylvania State University, University Park, Pennsylvania 16802, USA}
\author{Zhen Bi}
\email{zjb5184@psu.edu}
\affiliation{Department of Physics, The Pennsylvania State University, University Park, Pennsylvania 16802, USA}
\affiliation{Center for Theory of Emergent Quantum Matter, The Pennsylvania State University, University Park, Pennsylvania 16802, USA}

\begin{abstract}
Kagome systems are an ideal platform for exploring strongly correlated phases due to their unique electronic structure and geometric frustration. While recent solid-state realizations have uncovered a wealth of correlated states, they suffer from key limitations, including limited tunability of carrier density and interaction strength. Here, we propose an experimentally viable scheme to realize a breathing kagome moir\'e superlattice using a twisted trilayer of transition metal dichalcogenides (TMDs). By twisting the top and bottom layers relative to the middle by small angles $\theta$ and $2\theta$, respectively, we generate a kagome-like moir\'e potential on the central layer. Continuum model calculations reveal isolated kagome bands featuring flat bands, Dirac points with tunable gaps, and van Hove singularities. Crucially, this platform offers unprecedented control over band structure, carrier density, and interactions—achievable via twist angle and electrostatic gating. Our work opens a new route to realizing clean, tunable kagome metals and provides a versatile platform for studying strongly correlated and topological phenomena.
\end{abstract}

\maketitle

\section{Introduction} The kagome lattice\cite{Syozi1951} -- a two-dimensional network of corner-sharing triangles -- has been a focal point in the study of strongly correlated electronic states and exotic quantum phases of matter. Its unique geometry gives rise to a wealth of unconventional behaviors, including flat electronic bands, Dirac cones, and inherent geometric frustration. These characteristics make the kagome lattice an ideal platform to explore emergent states such as quantum spin liquids\cite{Sachdev1992,Yan2011,Balents2010,Broholm2020}, topologically nontrivial phases\cite{Guo2009,Tang2011,Xu2015}, and unconventional superconductivity\cite{Kiesel2013,Ko2009,Wang2013}. 

Cu-based compounds such as herbertsmithite, kapellasite, and haydeeite have been extensively studied for their potential quantum spin liquid behavior\cite{Normanreview2016,Shores2005,Helton2007,Mendels2007,Janson2008, Han2012, Imai2008, Okamoto2007}. More recently, the AV$_3$Sb$_5$ family (with A = Cs, Rb, K) has attracted attention for the coexistence of superconductivity and charge-density wave order\cite{Ortiz2019,Ortiz2020,Ortiz2021,Yin2021, Jiang2021,Li2021, Zhao2021, Chen2021}, while iron-based kagome compounds like FeSn and Fe$_3$Sn$_2$ are notable for their flat bands and Dirac fermion states\cite{Ye2018,Kang2020,Li2020FeSn}. Magnetic kagome systems, including Co$_3$Sn$_2$S$_2$, Mn$_3$Sn, and Mn$_3$Ge, have also been recognized for exhibiting large anomalous Hall effects and topological semimetallic behavior\cite{Kuroda2017,Nakatsuji2015,Nayak2016,Ikhlas2017}. However, these materials often face challenges such as structural distortions, chemical disorder, and -- more importantly -- a limited tunability of their electronic structure, charge density, and interaction strength. These motivate the search for new kagome systems with enhanced material quality and controllability.

Moir\'e materials\cite{Andrei2021,Moire2024} -- including twisted graphene systems\cite{Bistritzer2011} and transition metal dichalcogenides\cite{WuMacDonald2018,WuMacDonald2019} (TMDs) -- offer exceptional tunability in electronic structure and many-body correlations. Their success has revealed a wealth of strongly correlated phenomena, ranging from correlated insulating states\cite{Cao2018a, Chen2020CI,Tang2020,Regan2020, Wang2020CI,Li2021CI,Xu2020CI} and unconventional superconductivity\cite{Cao2018b, Chen2019SC,Liu2020mag,Park2021SC,Oh2021, Guo2024,Xia2024SC, Han2025} to integer and fractional anomalous quantum Hall effects\cite{Serlin2020,Li2021QAH,Cai2023,Park2023,Zeng2023,Lu2024,Foutty2024}. External knobs such as electrostatic gating, magnetic fields, pressure, and twist angle enable control of these systems, providing a versatile platform for engineering novel quantum phases. In this work, we demonstrate that twisted homotrilayer TMDs can be harnessed to realize kagome moir\'e lattices, potentially opening new avenues for investigating strongly correlated electrons on kagome lattices.


\section{Kagome Moiré superlattices}
\subsection{Twist scheme and moir\'e potential} 
\begin{figure}[h!]
\centering
\includegraphics[width=0.47\textwidth]{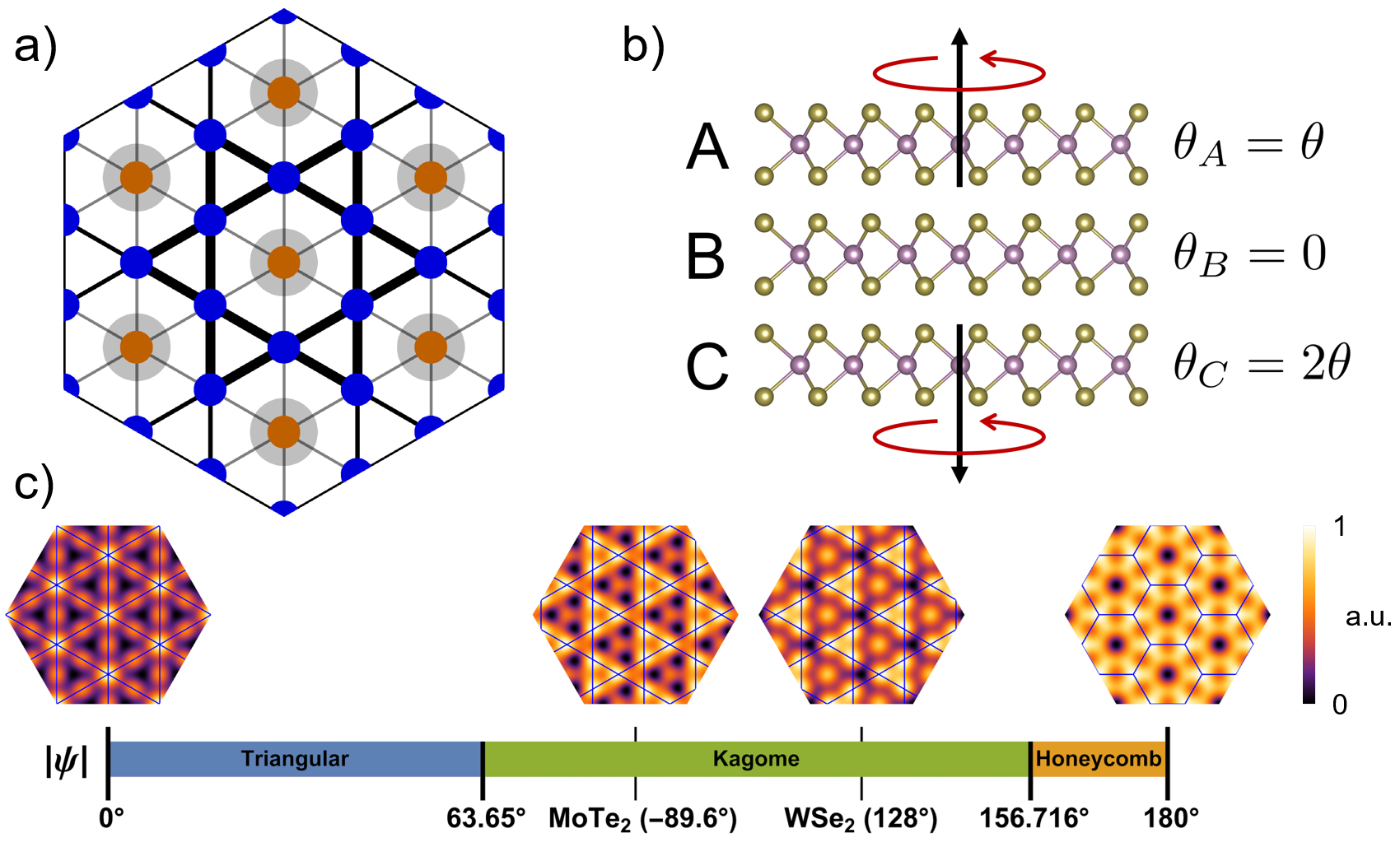}
\caption{\label{Fig1} (a) Obtaining a kagome lattice by deleting 1/4 of the sites from a triangular lattice; (b) The alternating twist schematic; (c) The total moir\'e potential experienced by layer B for different values of $\psi$. Color scales are independently normalized for each value of $\psi$ to emphasize spatial variations. Only relative extrema are physically relevant here. The maxima of the potentials form lattices that are indicated with blue lines. We observe that, as a function of the parameter $|\psi|$, there are three different regimes, namely triangular, kagome, and honeycomb. MoTe$_2$ \cite{WuMacDonald2019} and WSe$_2$\cite{Devakul2021} both fall in the kagome regime.}
\end{figure}

The twist scheme we designed stems from the observation that a kagome lattice can be formed by removing one quarter of the sites from a triangular lattice, where the removed sites lie on a larger triangular lattice with twice the original lattice constant, as shown in Fig. \ref{Fig1}(a). In practice, the initial triangular lattice is given by moir\'e potentials from twisted bilayers of TMD materials, while the larger triangular lattice for site removal can be achieved by twisting another TMD layer with half of the twist angle, and hence, twice the moir\'e lattice constant. The twist scheme is shown in Fig. \ref{Fig1}(b). We will show that overlaying these two moir\'e potentials is possible to generate a breathing kagome lattice potential on the middle layer. 


We begin by analyzing the moir\'e potentials within the continuum model for AA-stacked twisted TMD homobilayers\footnote{AA-stacking refers to the arrangement in which one monolayer is placed directly atop another, maintaining identical orientation and atomic positions in the plane. On the other hand, AB-stacking refers to the cases where the two layers are rotated by 180 degrees.}. Let us consider layer A (top) is twisted counterclockwise relative to layer B (bottom) by an angle $\theta$. In the simplest version of the continuum model for twisted TMDs -- where only the first harmonic potential is considered -- the moir\'e potential on each layer is modeled by the following form\cite{WuMacDonald2019}:
\begin{equation}\label{moire_potential_def}
    V_{\mp}^{AB}(\boldsymbol{r})=2V\sum_{j=1}^{3}\cos{\left(\boldsymbol{g}^{AB}_j\cdot\boldsymbol{r}\mp\psi\right)}
\end{equation}
where $\boldsymbol{g}_j^{AB}$'s are the reciprocal lattice vectors of the moir\'e lattice formed by layers A and B, and the $V$ and $\psi$ are material specific parameters. Here the $-/+$ in $V_{\mp}^{AB}(\boldsymbol{r})$ labels the potential on the top (A)/bottom (B) layer. The maxima of this potential form a triangular lattice for generic values of $V$ and $\psi$\footnote{Except at special values of $\psi$ such as $\psi=60^{\circ}$ with $V>0$, where the potential forms a honeycomb lattice.}. The position and shape of the maxima of the potential depend on the parameter $\psi$. 

Overlapping two moir\'e potentials to form a kagome potential can be done with three twisted TMD layers, since the middle layer will feel the moir\'e potentials from both sides. More explicitly, the potential that layer B experiences will be a combination of the AB and the BC moir\'e potentials.  If the lattice constant of one moir\'e potential is twice that of the other, the resulting maxima experienced by layer B can arrange into a kagome lattice. In this work, we consider the case where all three layers are composed of the same TMD material. We briefly comment on other cases in the discussion. 

To form a kagome potential, the layers must be twisted so that the lattice constant for the AB moir\'e lattice is twice that of the BC moir\'e lattice (i.e., $\left|a_M^{AB}/a_M^{BC}\right|=2$). Equivalently, the relative twist angles must satisfy $|\theta_{BC}/\theta_{AB}|=2$, where $\theta_{ll'}=\theta_l-\theta_{l'}$. In principle, one can implement either a helical twist ($\theta_{BC}=2\theta_{AB}$) or an alternating twist ($\theta_{BC}=-2\theta_{AB}$). In this work, we adopt the alternating-twist configuration by choosing $\theta_A=\theta$, $\theta_B=0$, and $\theta_C=2\theta$, which yields $\theta_{AB}=\theta$ and $\theta_{BC}=-2\theta$ (see Appendix A).

In the alternating twist case, the combined moir\'e potential felt by the middle layer can be written as 
\begin{align}
    \label{ABC_potential}\nonumber
    &V^{AB}_{+}(\boldsymbol{r})+V^{BC}_{-} (\boldsymbol{r}) \\ \nonumber
    &=2V\sum_{j=1}^3\left(\cos{\left(\boldsymbol{g}_j^{AB}\cdot\boldsymbol{r}+\psi\right)}+\cos{\left(\boldsymbol{g}_j^{BC}\cdot\boldsymbol{r}-\psi\right)}\right) \\    &=2V\sum_{j=1}^3\left(\cos{\left(\boldsymbol{g}_j^{AB}\cdot\boldsymbol{r}+\psi\right)}+\cos{\left(2\boldsymbol{g}_j^{AB}\cdot\boldsymbol{r}+\psi\right)}\right)
\end{align}
Here, from the second to the third line, we use the fact $\boldsymbol{g}_j^{BC}\cong-2\boldsymbol{g}_j^{AB}$ for the alternating twist. The shape of the moir\'e potential depends crucially on the parameter $\psi$. We treat $\psi$ as a tuning parameter and plot the moir\'e potential as a function of $\psi$ in Fig. \ref{Fig1}(c). We find three distinct regimes based on the number of potential maxima: for $0^{\circ} < \psi < 63.65^{\circ}$, the maxima form a triangular lattice. For $63.65^{\circ} < \psi < 156.716^{\circ}$, the maxima form a breathing kagome lattice. For $156.716^{\circ} < \psi < 180^{\circ}$, the maxima form a honeycomb lattice. The potentials for $-\psi$ are identical to those for $\psi$, except that they are rotated by $180^{\circ}$ in real space. Strikingly, a substantial portion of the parameter space yields a kagome-like lattice; notably, the parameters for both MoTe$_2$ and WSe$_2$ from Ref. \cite{WuMacDonald2019,Jia2024,Devakul2021,Xu2024,Reddy2023,Wang2024,Morales2023} fall within this regime.  

The moir\'e potential offers a preliminary guideline for identifying the regime where kagome physics might emerge. However, the realistic band structure is influenced by more than just the moir\'e potential on layer B, since layers A and C also play active roles, with significant tunneling to and from layer B. Therefore, to validate the proposed scheme, it is essential to perform a detailed continuum model calculation that incorporates all of these effects, which we will discuss next.

\begin{figure}[h!]
\centering
\includegraphics[width=0.45\textwidth]{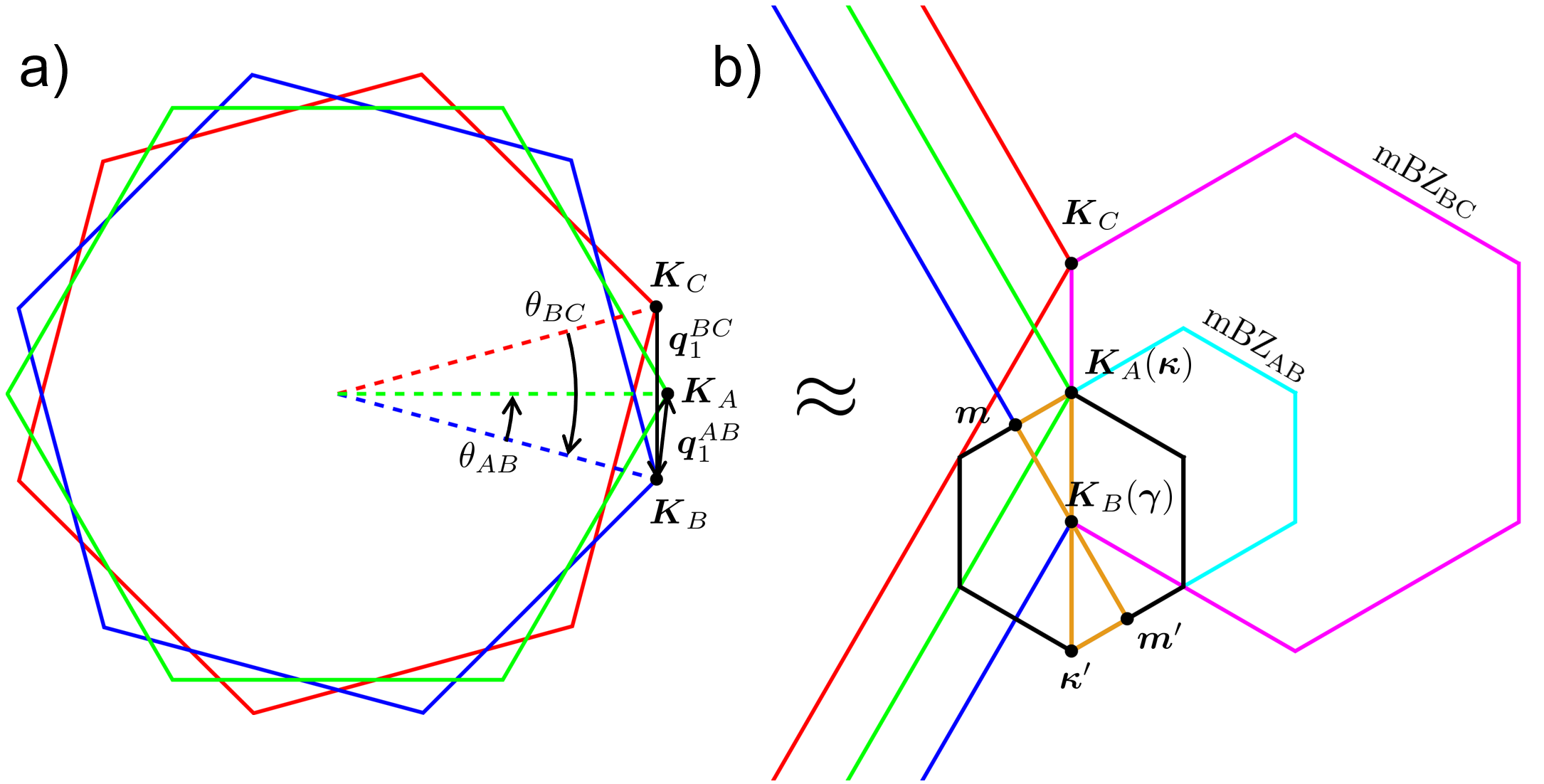}
\caption{\label{Fig2} (a) The Brillouin zones (BZs) associated with the alternating twist scheme. The BZs of the layers A, B, and C are green, blue, and red respectively. (b) The approximation we use in this work to calculate the band structure. The cyan BZ is the moir\'e BZ (mBZ) of layers A and B. The magenta BZ is the mBZ between layers B and C. The black BZ (just the cyan BZ but shifted) is the mBZ that is used in this paper. The yellow path is the path for band structure plots.}
\end{figure}

\subsection{Continuum model and band structures} Before proceeding with the calculations, one issue must be clarified. With the twist scheme described above, the orientations of the moir\'e superlattices generated by the AB and BC layers are slightly rotated relative to each other. This is evident in Fig.~\ref{Fig2}(a), where the edges ($\boldsymbol{q}^{AB}_1$ and $\boldsymbol{q}^{BC}_1$) of the AB and BC moir\'e Brillouin zones are slightly misaligned. This effect will lead to a supermoir\'e pattern. The lattice constant of this supermoir\'e pattern is on the order of $a^{AB,BC}_{MM} \sim a_0/\theta^2$,
which, for small twist angles, is much larger than those of the individual AB and BC moir\'e patterns. For instance, when the twist angle $\theta_{AB}$ is approximately $1.5^{\circ}$, the moir\'e lattice constant is around 10 nm, while the supermoir\'e lattice constant is on the order of 500 nm. Given that the electron coherence lengths in current TMD materials\cite{Neal2013,Qu2024} are much shorter than this, the supermoir\'e patterns can be safely neglected at this stage. Consequently, the band structure can be determined using only the AB and BC moir\'e patterns, assuming they are aligned. Although the band structure would exhibit variations on the scale of the supermoir\'e lattice constant, we defer the study of these variations to future work. In summary, we approximate the AB and BC moir\'e lattices as being commensurate (with collinear $\boldsymbol{K}$-points), and the AB moir\'e Brillouin zones will serve as the effective Brillouin zone as shown in Fig. \ref{Fig2}(b).

The AA-stacked twisted homotrilayer TMD continuum model for valley $\eta=\pm$ is given by the Eq. (\ref{homotrilayer_hamiltonian}).
\begin{widetext}
\begin{equation}\label{homotrilayer_hamiltonian}
    H_{\eta}(\boldsymbol{r})=
    \begin{pmatrix}
        -\frac{\hbar^2\hat{\boldsymbol{k}}^2}{2m^*}+V^{AB}_{-}(\boldsymbol{r})+D/2&t^{AB}_{\eta}(\boldsymbol{r})&0\\
        t^{AB*}_{\eta}(\boldsymbol{r})&-\frac{\hbar^2\hat{\boldsymbol{k}}^2}{2m^*}+V^{AB}_{+}(\boldsymbol{r})+V^{BC}_{-}(\boldsymbol{r})&t^{BC}_{\eta}(\boldsymbol{r})\\
        0&t^{BC*}_{\eta}(\boldsymbol{r})&-\frac{\hbar^2\hat{\boldsymbol{k}}^2}{2m^*}+V^{BC}_{+}(\boldsymbol{r})-D/2
    \end{pmatrix}
\end{equation}
\end{widetext}
The form of the moir\'e potentials are given by Eq. (\ref{moire_potential_def}) and (\ref{ABC_potential}). The inter-layer tunneling is given by
\begin{equation}\label{interlayer_tunneling_def}
    t^{ll'}_{\eta}(\boldsymbol{r})=w\sum_{j=1}^{3}{e^{-\eta\, i\,\boldsymbol{q}_j^{ll'}\cdot\boldsymbol{r}}},
\end{equation}
where the tunneling strength $w$ is material dependent, $l$ and $l'$ label layers. Here, the moir\'e scattering momentum is $\boldsymbol{q}_1^{ll'}=\boldsymbol{K}_{l}-\boldsymbol{K}_{l'}$ and rotations of $\boldsymbol{q}_1^{ll'}$ give $\boldsymbol{q}_j^{ll'}=\mathcal{C}_3^{j-1}\boldsymbol{q}_1^{ll'}$. The moir\'e reciprocal lattice vectors $\boldsymbol{g}^{ll'}_j$ are given by $\boldsymbol{g}^{ll'}_j=\mathcal{C}_3^{j-1}\boldsymbol{g}^{ll'}_1=\boldsymbol{q}_{j+2}^{ll'}-\boldsymbol{q}_{j+1}^{ll'}$. $D$ is the vertical displacement field which is an external parameter. The degrees of freedom in the two valleys are related by time reversal symmetry. 

\begin{figure}[h!]
\centering
\includegraphics[width=0.45\textwidth]{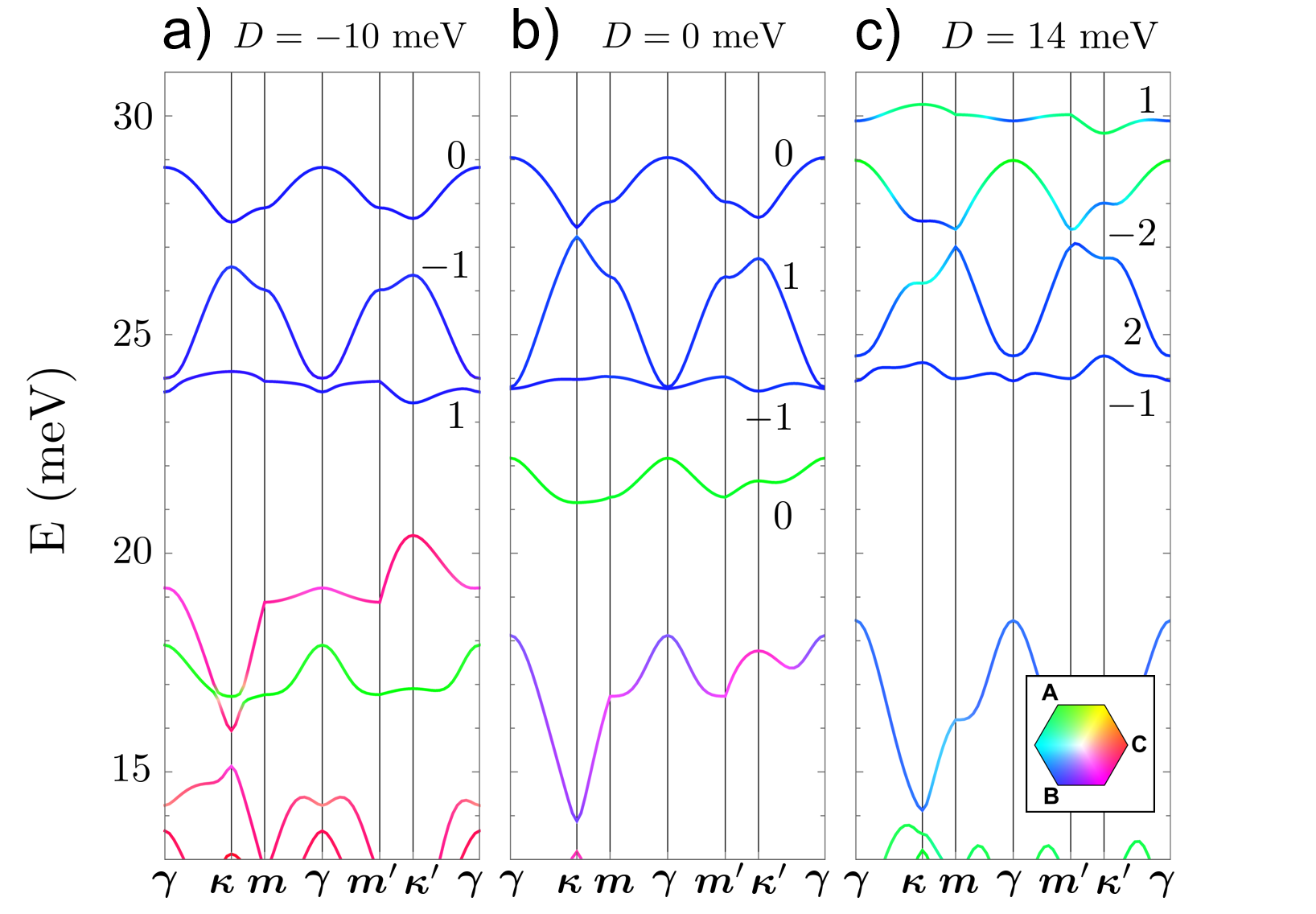}
\caption{\label{Fig3} Band structure for MoTe$_2$ at $\theta=1.2^{\circ}$ for displacement fields at (a) -10 meV, (b) 0 meV, and (c) 14 meV. The colors of the bands indicate the layer polarizations. The Chern numbers of the top few bands are labeled on the side.}
\end{figure}

The parameters used for MoTe$_2$\cite{WuMacDonald2019} are $V=8$ meV, $w=-8.5$ meV, $\psi=-89.6^{\circ}$, $a=3.472$ \AA, and $m^*=0.62\,m_e$, where $m_e$ is the electron mass. We plot the band structure from the continuum model for $\theta\equiv\theta_{AB}=1.2^{\circ}$ in Fig. \ref{Fig3}. The path in $k$-space is shown in yellow in Fig. \ref{Fig2}(b).  With the MoTe$_2$ parameters, we observe discernible kagome band structures up to $\theta\sim 1.5^\circ$. We also analyze the band structure with WSe$_2$ parameters; the results are detailed in the Appendix B.

The moir\'e kagome lattice realized in our construction has lower symmetry than an ideal kagome lattice. Within the commensurate continuum description, the single-valley Hamiltonian is invariant under translations of the enlarged moir\'e unit cell and under the threefold rotation $C_{3z}$. In contrast to the $D_{6h}$ point-group symmetry of an ideal kagome lattice, the alternating-twist geometry distinguishes the right- and left-pointing triangles and generically breaks sixfold rotation, inversion, and unitary mirror symmetries. The resulting structure is therefore a breathing kagome lattice with only exact $C_3$ point-group symmetry. In particular, the two inequivalent corners $\kappa$ and $\kappa'$ of the moir\'e Brillouin zone are not required to be degenerate. The complete two-valley Hamiltonian additionally preserves microscopic time-reversal symmetry $\mathcal{T}$, which exchanges the two valleys and relates their Berry curvatures and Chern numbers. We emphasize that moir\'e translation symmetry is exact only within the commensurate approximation adopted here; in the physical trilayer it is weakly modulated on the much longer supermoir\'e length scale.

Nevertheless, the top three bands of this moir\'e system strongly resemble a kagome band structure. The layer polarization of the Bloch wavefunctions is depicted using RGB colors, with the layers A, B, and C represented by green, blue, and red, respectively. As shown in Fig. \ref{Fig3}, these top three bands are almost entirely localized on layer B and are energetically well separated from the rest of the moir\'e bands without any additional tuning.


A comment on the band ordering is in order. For electrons in a conventional kagome confining potential, the kagome flat band is expected to appear at the \emph{top} of the relevant three-band manifold. In our continuum Hamiltonian Eq.~\eqref{homotrilayer_hamiltonian}, however, the kinetic term $-\hbar^{2}\hat{\boldsymbol{k}}^{2}/2m^{*}$ describes \emph{holes} in the TMD valence band, for which the moir\'e potential maxima act as confining wells. The resulting particle-hole conjugation reverses the band ordering, placing the kagome flat band at the \emph{bottom} of the upper three-band manifold. Reaching the flat band therefore requires populating the two dispersive kagome bands above it, corresponding to a moderate hole doping that is well within the range routinely accessible by electrostatic gating in moir\'e TMD devices.

However, we emphasize that tunneling between layers A and B actually brings the system closer to the ideal kagome band structure. If the tunneling between layers B and C is turned off (\(t^{BC}_{\eta}(\boldsymbol{r})=0\)), the band structure still resembles kagome bands. In contrast, if the tunneling between A and B layers is turned off (\(t^{AB}_{\eta}(\boldsymbol{r})=0\)), the band gap at the $\boldsymbol{\kappa}$-points becomes much larger, indicating a much more pronounced “breathing” effect (see Appendix B for details).

The fourth band is almost completely localized on layer A. This band can be shifted in energy by tuning the vertical displacement field $D$. In particular, to isolate the kagome bands, one can apply a displacement field to push the layer A band away (see Fig.~\ref{Fig3}(a)). Alternatively, raising the layer A band to interact with the kagome bands can modulate the topology of the kagome bands, as illustrated in Fig.~\ref{Fig3}(c). This additional degree of freedom adds further complexity and tunability to the model.

The Chern numbers shown in Fig.~\ref{Fig3} reveal a non-trivial $D$-dependence. At $D = -10$~meV the layer-A orbital is energetically separated from the kagome manifold and the top three bands carry Chern numbers $(0, -1, +1)$ from top
to bottom, consistent with a breathing-kagome model with complex NN hoppings whose phases generate a kagome analog of the Haldane mass. As $D$ is increased, the layer-A band is
pushed upward and eventually hybridizes with the topmost kagome band near the $\gamma$ point. This hybridization inverts the band character and redistributes Chern numbers among the bands; at $D = +14$~meV Fig.~\ref{Fig3}(c) the top four bands carry $(1, -2, 2, -1)$. The displacement field thus acts as a clean experimental knob for driving topological transitions within the kagome manifold by exchanging Berry curvature with the layer-A ancilla orbital, a feature absent in other kagome moiré platforms including the graphene-based kagome moir\'e proposals of Refs.~\onlinecite{Scheer2022,Scheer2023}.

\begin{figure}[h!]
\centering
\includegraphics[width=0.39\textwidth]{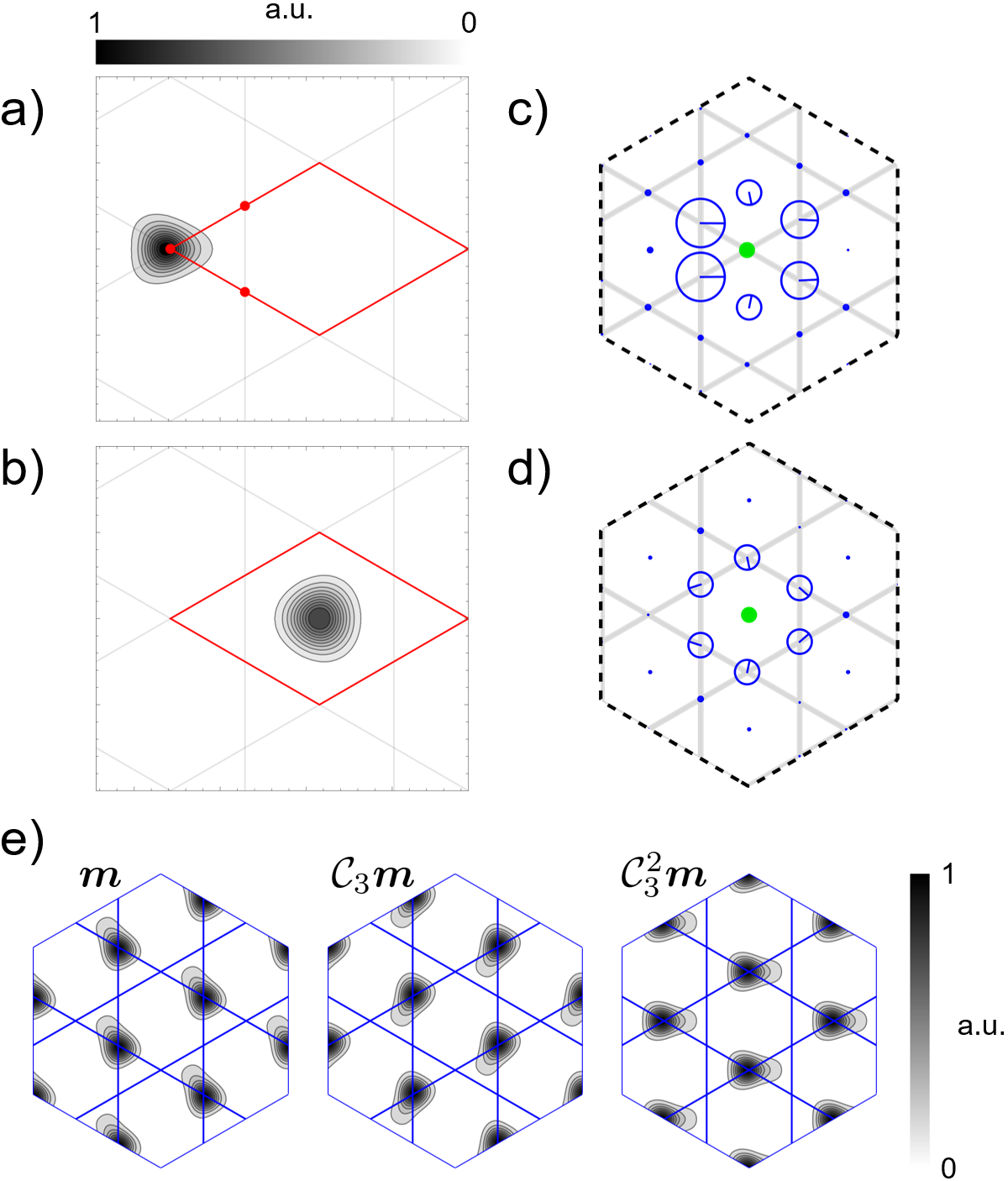}
\caption{\label{Fig4}  (a) (b) The total density of the Wannier functions of MoTe$_2$ with $\theta=1.2^\circ$. The red lines outline the moir\'e unit cell. An illustration for the hopping parameters from a kagome site (c) and from a site on layer A (d). For each hopping from the center green orbital, the radius of the circle is proportional to the amplitude of that hopping, and the radius shown is in the direction of the phase. (e) Sublattice polarizations of the wavefunctions for points $\boldsymbol{m}$, $\mathcal{C}_3\boldsymbol{m}$, and $\mathcal{C}_3^2\boldsymbol{m}$ at the van Hove singularity points on the second band from the top.}
\end{figure}

\subsection{Wannier functions and tight-binding models}
\begin{figure}[t]
\centering
\includegraphics[width=0.8\linewidth]{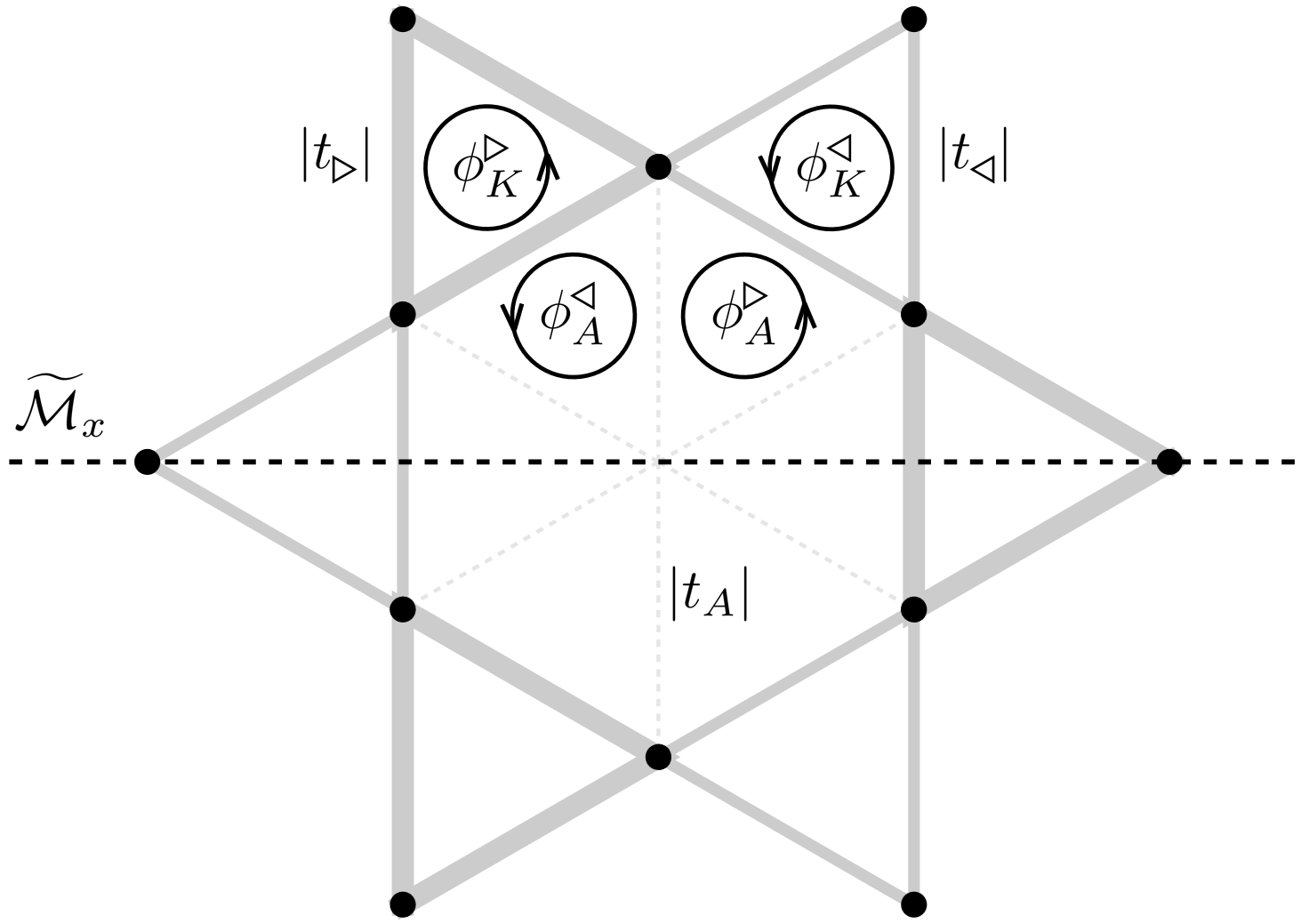}
\caption{\label{fluxfig}The symmetry-inequivalent triangular plaquettes of twisted trilayer MoTe$_2$ are labeled and given a counterclockwise orientation. The acquired fluxes and the magnitudes of the hopping amplitudes as a function of twist angle $\theta$ and as a function of displacement field $D$ are given in Tables~\ref{fluxtable} and \ref{fluxtableD} respectively in the Appendix. The thicker bonds indicate larger-magnitude hopping amplitudes for the right-pointing kagome triangles than for the left-pointing kagome triangles. For twisted trilayer MoTe$_2$ with $\theta=1.2^\circ$, the acquired fluxes are $\phi_K^{\triangleright}=-3.7^\circ$, $\phi_K^{\triangleleft}=5.7^\circ$, $\phi_A^{\triangleright}=34.7^\circ$, and $\phi_A^{\triangleleft}=84.6^\circ$, while the magnitudes of the hopping amplitudes are $|t_{\triangleright}|=1.055$ meV, $|t_{\triangleleft}|=0.800$ meV, and $|t_A|=0.533$ meV. The dashed line labeled $\widetilde{\mathcal{M}}_x$ denotes the mirror plane associated with the approximate emergent magnetic-mirror symmetry.}
\end{figure}

We obtain the Wannier functions for the top four bands of MoTe$_2$ with $\theta=1.2^\circ$ using Wannier90\cite{Pizzi2020} and construct the corresponding tight-binding model by extracting the real-space Hamiltonian matrix elements in the Wannier basis. The first three maximally localized Wannier functions (MLWFs) are localized at the sites of the kagome lattice on layer B, as expected; the total density of one of these functions is shown in Fig.~\ref{Fig4}(a), while the other two are obtained by rotations of the MLWF about the origin by \(2\pi/3\) and \(4\pi/3\), taking it to the other two red points in Fig.~\ref{Fig4}(a). The fourth MLWF is centered at the maxima on layer A, and its total density is depicted in Fig.~\ref{Fig4}(b). The hopping parameters decay exponentially with distance. A graphical representation of the hopping parameters for the first MLWF is shown in Fig.~\ref{Fig4}(c), and one for the fourth MLWF is shown in Fig.~\ref{Fig4}(d). Restricting to nearest-neighbor hopping, symmetry divides the bonds into three classes \(T\), as indicated in Fig.~\ref{fluxfig}. The hopping amplitude on a bond \(\langle ij\rangle \in T\) can be parameterized as \(|t_T|e^{i\theta_{ij}}\). Although the individual bond phases are gauge dependent, the accumulated phase around a closed loop is gauge invariant. We therefore define four gauge-invariant fluxes associated with the four triangular plaquettes shown in Fig.~\ref{fluxfig}. In general, we find nonzero fluxes through all four plaquettes, which can be tuned smoothly with the twist angle and displacement field. Representative values are shown in Tables~\ref{fluxtable} and \ref{fluxtableD} in the Appendix. Despite the tunability, the total flux through each unit cell is \(2\pi\). More details of the MLWFs and hopping data are provided in Appendix C.

Although the twisted trilayer lacks a microscopic mirror symmetry, its low-energy tight-binding model exhibits an approximate emergent magnetic-mirror symmetry $\widetilde{\mathcal{M}}_x=\mathcal{M}_x\mathcal{T}$, where $\mathcal{M}_x$ denotes reflection about the $x$ axis as shown in Fig.~\ref{fluxfig}. Although $\mathcal{M}_x$ and time reversal $\mathcal{T}$ separately exchange the two valleys, their product preserves each valley, maps $(k_x,k_y)$ to $(-k_x,k_y)$, and satisfies $\widetilde{\mathcal{M}}_x^2=+1$ for spinful Kramers valleys. Without imposing this symmetry in the Wannier construction, we find that the resulting band structure approximately satisfies $E_n(k_x,k_y)=E_n(-k_x,k_y)$ and that the fitted hopping amplitudes approximately obey the corresponding magnetic-mirror relations. We therefore regard $\widetilde{\mathcal{M}}_x$ not as an exact microscopic symmetry of the twisted trilayer, but as an approximate emergent symmetry of its low-energy moir\'e description.

An interesting feature to note is the sublattice polarization of this kagome model. We plot the real-space distribution of the Bloch wavefunctions near the van Hove singularity (VHS), namely $\boldsymbol{m}$-points, for the top three bands in Appendix D. We find that the real-space wavefunction of each $\boldsymbol{m}$-point in the second band is localized on one sublattice of the kagome lattice (shown in Fig. \ref{Fig4}(e)). The pure-sublattice character of the VHS on the middle band is particularly interesting. In the standard sublattice-interference scenario for the kagome
Hubbard model, a sublattice polarized VHS suppresses the on-site bare susceptibility while enhancing nesting-driven instabilities in non-trivial angular-momentum channels. At the corresponding van Hove filling ($n \cong 5/12$ of the kagome manifold), repulsive interactions are predicted to drive instabilities toward $f$-wave and chiral $d \pm i d$ superconductivity, as well as toward bond-ordered density waves and orbital-current states~\cite{Kiesel2012,Wu2023,Wang2013VanHove}. In our platform, this filling is reached at a hole density of $n_{h} \cong 6\times (5/12)/A_{M}$ per moir\'e unit cell, where
$A_{M} = (\sqrt{3}/2) a_{M}^{2}$ is the moir\'e unit cell area. For $\theta = 1.2^{\circ}$ in MoTe$_{2}$, this gives $a_{M} \simeq 16.6~\mathrm{nm}$, $A_{M} \simeq 238~\mathrm{nm}^{2}$, and $n_{h} \simeq 1.0 \times 10^{12}~\mathrm{cm}^{-2}$, well within the range routinely accessible by electrostatic gating in moir\'e TMD devices.



\section{Tunable spin models at half-filling} 

While our analysis so far has focused on the noninteracting band structure, it is important to understand how electronic interactions enter in a controlled limit. We therefore consider a Hubbard interaction at half filling and analyze the large-$U$ regime using degenerate perturbation theory.

Starting from
\begin{equation}
    H = \sum_i \varepsilon_i n_i + \sum_i U_i\, n_{i\uparrow} n_{i\downarrow} + V ,
\end{equation}
where $V$ contains the hopping processes introduced in the noninteracting model, charge fluctuations are suppressed for $U_i \gg |t|$, and the low-energy Hilbert space consists of singly occupied states. A second-order expansion in $t/U$ yields an effective spin Hamiltonian acting within this manifold (see Appendix E).

For time-reversal-symmetric hopping amplitudes, the resulting effective Hamiltonian can be written as
\begin{align}\label{Heff_final}
    H_{\mathrm{eff}}=\sum_{T}\frac{4 |t_T|^2}{U_T}
    \sum_{\langle ij \rangle \in T}
    &\left[S_i^z S_j^z\right. \nonumber\\
        &+ \cos(2\theta_{ij})
        \left(S_i^x S_j^x + S_i^y S_j^y \right) \nonumber \\
        &+\left. \sin(2\theta_{ij})
        (\vec{S}_i \times \vec{S}_j )_z\right],
\end{align}
where $T$ labels the symmetry-inequivalent bond classes of the lattice and $\theta_{ij}$ are the bond phases. The parameters $U_T$ denote the effective interaction strengths associated with each bond class and are determined by the onsite energies and Hubbard interactions of the two sites connected by the bond. As an example, for a bond $\braket{ij}\in T$, we have $U_T=U_{ij}=2((U_i+\varepsilon_i-\varepsilon_j)^{-1}+(U_j+\varepsilon_j-\varepsilon_i)^{-1})^{-1}$.

For the breathing kagome lattice with time-reversal symmetry, the lattice contains two symmetry-inequivalent nearest-neighbor kagome bonds, those belonging to right-pointing and left-pointing triangles.
The onsite interaction on kagome sites is uniform, so that $U_{\triright}=U_{\trileft}$ along with $\varepsilon_{\triright}=\varepsilon_{\trileft}$, while the corresponding hopping amplitudes are characterized by two independent complex parameters $t_{\triright}$ and $t_{\trileft}$.

Introducing additional sites at the centers of the kagome hexagons adds a further onsite interaction parameter $U_{\hexcenterdot}$.
With only $\mathcal{C}_3$ rotational symmetry, the bonds connecting hexagon-center sites to kagome sites fall into two symmetry-inequivalent classes, described by the complex hopping amplitudes $t_{\trilinesU}$ and $t_{\trilinesD}$, as indicated by their subscripts. Since these bonds connect kagome sites to hexagon-center sites, the corresponding effective interaction strengths are given by
\begin{equation*}
    U_{\trilinesU} = U_{\trilinesD} = 2\left((U_{\triright}+\varepsilon_{\triright}-\varepsilon_{\hexcenterdot})^{-1} + (U_{\hexcenterdot}+\varepsilon_{\hexcenterdot}-\varepsilon_{\triright})^{-1}\right)^{-1}.
\end{equation*}
The sum over bond classes in Eq. (\ref{Heff_final}) therefore runs over
$T = \triright,\, \trileft,\, \trilinesU,\, \trilinesD$. With the approximate emergent magnetic-mirror symmetry $\widetilde{\mathcal{M}}_x$, the hopping amplitudes along $\trilinesU$ and $\trilinesD$ are related by complex conjugation. This symmetry therefore combines the two bond classes into a single class, $\hexlines$.

This provides a controlled interacting extension of the noninteracting kagome plus hexagon center band structure discussed above. The onsite energy of the spin at the hexagon center is directly controlled by the vertical displacement field since it is localized on layer A, so the system can be continuously tuned between a pure kagome spin model (central spin pushed far above or below the kagome manifold) and a coupled kagome--triangular spin system (central spin resonant with the kagome sites). This provides a clean experimental knob to interpolate between qualitatively distinct correlated regimes within a single device. 

While a complete phase-diagram analysis of Eq.~\eqref{Heff_final} is beyond the scope of this work, several limiting regimes are already intriguing.

\emph{(i) Pure breathing-kagome limit
($|\varepsilon_{\hexcenterdot}| \to \infty$).} When the displacement field pushes the layer-A orbital far from the kagome manifold, the central spin decouples and Eq.~\eqref{Heff_final} reduces to a breathing-kagome XXZ + Dzyaloshinskii-Moriya (DM) interaction model with two independent triangular exchange couplings $J_{\triright}$ and $J_{\trileft}$ and DM vectors $\boldsymbol{D}_{\triright}$, $\boldsymbol{D}_{\trileft}$ pinned
along $\hat{\boldsymbol{z}}$ by $C_{3}$ symmetry. 
(Note that the DM interaction in our model cannot be globally gauged away by a local rotation of the spin basis, because the plaquettes carry nontrivial gauge-invariant fluxes. These fluxes can be continuously tuned by varying the twist angle and displacement field.) The ground state of the breathing-kagome Heisenberg model is the subject of active investigation: variational Monte Carlo studies have proposed both a gapped $\mathbb{Z}_{2}$ spin liquid~\cite{Schaffer2017,IqbalPoilblancSchuch2020} and a
persistent gapless U(1) Dirac spin liquid~\cite{IqbalPoilblancThomaleBecca2018}, while DMRG and
iPEPS calculations find that the isotropic spin liquid is
stable to substantial breathing anisotropy, with a transition to a nematic phase at small
$J_{\trileft}/J_{\triright}$~\cite{RepellinPollmannHe2017,JahromiOrusPoilblancMila2020}. The DM interactions present in our microscopic derivation provide an additional axis of the phase diagram that is relevant to the chiral spin liquid scenario~\cite{Messio2010}. The proposed platform thus provides a clean experimental setting for testing these competing theoretical predictions. 

\emph{(ii) Resonant central spin
($\varepsilon_{\hexcenterdot} \simeq \varepsilon_{\triright}$).} When
the layer-A orbital is tuned into resonance with the kagome
manifold, the central spin couples antiferromagnetically to
all six surrounding kagome sites, realizing a two-dimensional
spin model in which a kagome antiferromagnet is
augmented by an additional spin at every hexagon center.
Closely related models have been studied previously: the
spin-1/2 Heisenberg model interpolating between kagome and
triangular limits via a centered spin retains
120$^{\circ}$ Néel order down to $J'/J \simeq 0.2$, below
which antiferromagnetic correlations rapidly
weaken~\cite{ArracheaCapriottiSorella2004}; and a classical
Heisenberg model with a six-coordinated central moment
coupled to surrounding kagome spins selects the
$\sqrt{3}\times\sqrt{3}$ ordering over the $q=0$ ordering
at weak to moderate coupling, with crossovers to canted and
ferrimagnetic phases at stronger coupling~\cite{Redpath2012}.
Our platform realizes this class of models,
with the central-coupling strength continuously tunable by
the displacement field, providing a direct experimental
testbed for these predictions.

\emph{(iii) Doping away from half filling.}
Light doping of the half-filled Mott insulator gives rise to a breathing-kagome \(t\)-\(J\) model with complex hopping amplitudes. Previous studies of the uniform kagome \(t\)-\(J\) model have proposed unconventional superconducting states~\cite{KoLeeWen2009,JiangYaoYang2021}, and other competing states such as valence-bond and charge-ordered states~\cite{GuertlerMonien2013,JiangDevereauxKivelson2017}. The combination of breathing anisotropy and nontrivial hopping fluxes further enriches this already subtle many-body problem, opening new opportunities for pairing and competing correlated phases beyond those of the uniform kagome model. A quantitative numerical investigation of this regime is left to future work.

A quantitative numerical analysis of these regimes is left to future work.

\section{Discussion and outlook} In this work, we propose a route to achieve a kagome moir\'e lattice by twisting three layers of TMD materials. Our continuum model calculations exhibit a strong resemblance to the kagome band structure under this scheme. However, there are two caveats to note: 1) Our model is based on the first-harmonic continuum approximation, which is expected to perform well for intermediate twist angles. For smaller twist angles, lattice relaxation effects become significant, and it is argued that the continuum model needs to be updated to include higher-order harmonics\cite{Jia2024,Mao2024}. The exact models for small angles are currently under investigation, and the results have not yet converged\cite{Jia2024,Mao2024}. Consequently, we leave this aspect for future study. 2) As mentioned earlier, this particular twist scheme gives rise to a supermoir\'e lattice with a characteristic length scale of hundreds of nanometers. This implies that the continuum model band structure is valid only within regions of this size, and the band structure will vary on the scale of the supermoir\'e lattice. We emphasize that the validity of the local moir\'e band
description depends on which probe is being considered. For
transport experiments and any quantity dominated by the
single-particle coherence length
$\ell_{\rm coh} \ll a_{MM}$, the supermoir\'e variation
averages out and the local band structure controls the
response. For real-space probes such as STM and nano-ARPES,
however, the band structure should be expected to vary
continuously across the supermoir\'e unit cell, in analogy with
the moir\'e mosaic observed in helical trilayer
graphene~\cite{Devakul2023,Xia2025}. Understanding the impact of this mosaic physics for the kagome case is an interesting future direction.

We also note a few promising proposals for graphene-based kagome moir\'e systems in Refs. \onlinecite{Scheer2022,Scheer2023} and TMD-based kagome moiré systems in Ref. \onlinecite{Reddy2023k}, although they require a certain degree of parameter tuning or specific filling and interaction effects.
Another related approach to realizing kagome moir\'e superlattices in twisted TMDs was recently proposed in Ref. \onlinecite{Nakatsuji2025}. While this work also demonstrates the emergence of kagome moir\'e superlattices, the underlying mechanisms differ. Their construction relies on a lateral displacement between same-angle, helically twisted layers, whereas our construction employs alternating twist angles without any relative shift. As a result, the band structures and tunability are distinct. For example, the tunable fourth band from the top that is localized on layer A is unique to our approach.
We also emphasize that our approach may offer a general route to engineering line-graph moiré superlattices. The kagome lattice is the line graph of the honeycomb lattice, and our construction realizes it from a triangular moiré potential by selectively suppressing one quarter of the maxima with a commensurate longer-wavelength potential. The same principle could generalize to other parent lattices: applying the alternating-twist scheme to a square-lattice material, for instance, may have the potential to realize a Lieb-lattice moiré, with the longer-wavelength potential suppressing the appropriate sublattice of the parent square moiré. We leave a detailed exploration of such generalizations to future work.

Now we turn to whether trilayers composed of different TMD materials can be used to realize a kagome lattice. Unfortunately, in general, this approach does not work. The desired 1:2 ratio of moir\'e lattice constants can always be satisfied by a combination of mismatch and twist. However, once a lattice mismatch is present, the orientation of the resulting moir\'e patterns becomes highly sensitive to the mismatch value and the twist angle. Even if the 1:2 lattice constant ratio is achieved, the two moir\'e patterns are typically misaligned by a significant angle, which does not produce a kagome lattice. Nevertheless, by exploring the extensive library of TMD materials, one may identify cases where the lattice mismatches alone can yield moir\'e lattice constants that are approximately in a 1:2 ratio. For instance, consider a trilayer system composed of WTe$_2$, WSe$_2$, and MoS$_2$, with lattice constants 3.55\r{A}\cite{Zhao2013}, 3.28\r{A}\cite{Zhao2013}, and 3.16\r{A}\cite{Mak2010}, respectively. Assuming perfect alignment of their crystal orientations, the moir\'e pattern generated by WTe$_2$ and WSe$_2$ has a lattice constant of 4.67nm, while that from WSe$_2$ and MoS$_2$ has a lattice constant of 8.96nm, resulting in a roughly 1:2 ratio. One might then expect a similar kagome band structure in this case. A survey of available TMD materials may reveal more examples where a kagome moir\'e lattice could be viable. Of course, a detailed study of the band structure for these cases would require an advanced continuum model or large-scale DFT calculations, which we leave for future work.

\begin{acknowledgments}
\textit{Acknowledgements} -- We thank Ziqiang Wang, Yang Zhang, Jian-Hao Zhang, Yunzhe Liu, Binghai Yan, Chao-Ming Jian, Peizhe Tang, Xiaodong Xu, Ming Yi, Emanuel Tutuc, and Allan MacDonald for stimulating discussions. AF and ZB acknowledge the support from NSF through Grant DMR-2339319. AF is also supported by a GFSD Fellowship. CXL acknowledges the support from NSF through The Pennsylvania State University Materials Research Science and Engineering Center [DMR-2011839].
\end{acknowledgments} 

\appendix

\onecolumngrid

\section{Continuum model for twisted TMD}\label{app:continuum model}
The AA-stacked TMD homobilayer continuum model Hamiltonian between two adjacent layers (A and B) is given by Eq. (\ref{Ham_TMD_bilayer})\cite{WuMacDonald2019}, where A is on top of B and rotated by a small angle $\theta_{AB}$ counterclockwise from B, and $\eta=\pm$ labels the valley.
\begin{equation}\label{Ham_TMD_bilayer}
    H^{AB}_{\eta}(\boldsymbol{r})=
    \begin{pmatrix}
        -\frac{\hbar^2\hat{\boldsymbol{k}}^2}{2m^*}+V^{AB}_{-}(\boldsymbol{r})&t^{AB}_{\eta}(\boldsymbol{r})\\
        t^{AB*}_{\eta}(\boldsymbol{r})&-\frac{\hbar^2\hat{\boldsymbol{k}}^2}{2m^*}+V^{AB}_{+}(\boldsymbol{r})
    \end{pmatrix}
\end{equation}

\begin{figure}[h!]
\centering
\includegraphics[width=0.8\textwidth]{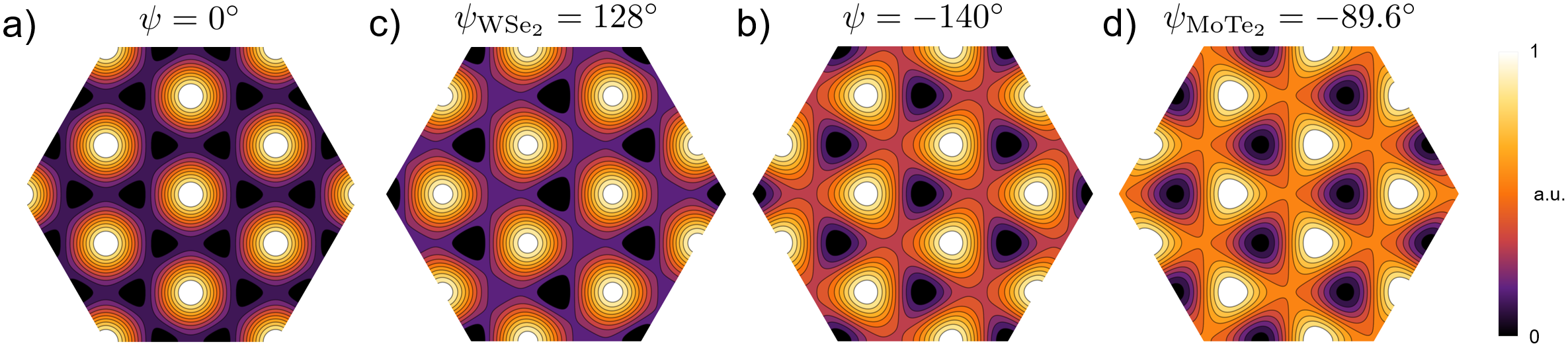}
\caption{\label{FigPotentials} The moir\'e potential on layer B for a bilayer system is plotted in real space for different values of the parameter $\psi$, including that of MoTe$_2$ \cite{WuMacDonald2019} and WSe$_2$\cite{Devakul2021}. The degenerate maxima of the potential form a triangular lattice. Color scales are independently normalized for each value of $\psi$ to emphasize spatial variations.}
\end{figure}

The moir\'e potential from layers A and B is approximated by including just the first harmonic of its Fourier expansion, see Eq. (\ref{moire_potential_def_app}). The potentials on the top (A) and on the bottom (B) layer are given by $V^{AB}_{-}(\boldsymbol{r})$ and $V^{AB}_{+}(\boldsymbol{r})$ respectively. 
\begin{equation}\label{moire_potential_def_app}
    V^{AB}_{\mp}(\boldsymbol{r})=2V\sum_{j=1}^{3}\cos{\left(\boldsymbol{g}_j^{AB}\cdot\boldsymbol{r}\mp\psi\right)}
\end{equation}
The amplitude $V$ and phase $\psi$ are parameters that are material specific. The moir\'e reciprocal lattice vectors $\boldsymbol{g}^{AB}_j$ are given by $\boldsymbol{g}^{AB}_j=\mathcal{C}_3^{j-1}\boldsymbol{g}^{AB}_1=\boldsymbol{q}_{j+2}^{AB}-\boldsymbol{q}_{j+1}^{AB}$ where $\boldsymbol{q}_1^{AB}=\boldsymbol{K}_{A}-\boldsymbol{K}_{B}$, and $\boldsymbol{q}_j^{AB}=\mathcal{C}_3^{j-1}\boldsymbol{q}_1^{AB}$. The potential respects $C_3$ symmetry. The positions and shape of maxima of the potential depend on the parameter $\psi$. We plot a few examples of the potentials for different values of $\psi$ in Fig. \ref{FigPotentials}.

The inter-layer tunneling amplitude between layers A and B is given by Eq. (\ref{interlayer_tunneling_def_app}), where the parameter $w$ is known as the tunneling strength. 
\begin{equation}\label{interlayer_tunneling_def_app}
    t^{AB}_{\eta}(\boldsymbol{r})=w\sum_{j=1}^{3}{e^{-\eta\, i\,\boldsymbol{q}_j^{AB}\cdot\boldsymbol{r}}}
\end{equation}

 The Hamiltonian in Ref. \cite{WuMacDonald2019} is related to the Hamiltonian in Eq. (\ref{Ham_TMD_bilayer}) by a unitary transformation $\widetilde{H}^{AB}_{\eta}(\boldsymbol{r})=U_\eta(\boldsymbol{r})H^{AB}_{\eta}(\boldsymbol{r})U^\dag_\eta(\boldsymbol{r})$ where $U_\eta(\boldsymbol{r})$ is given by Eq. (\ref{homobilayer_unitary}). The benefit of the form in Eq. (\ref{interlayer_tunneling_def_app}) is that the $\mathcal{C}_3$ symmetry is manifest since $H^{AB}_{\eta}(\mathcal{C}_3\boldsymbol{r})=H^{AB}_{\eta}(\boldsymbol{r})$.

\begin{equation}\label{homobilayer_unitary}
    U_\eta(\boldsymbol{r})=
    \begin{pmatrix}
        e^{\eta i\boldsymbol{K}_A\cdot\boldsymbol{r}}&0\\
        0&e^{\eta i\boldsymbol{K}_B\cdot\boldsymbol{r}}
    \end{pmatrix}
\end{equation}

For three twisted layers, the moir\'e potential on layer B is given by Eq. (\ref{B_potential}) since layer B is the bottom layer in the AB moir\'e system and the top layer in the BC moir\'e system.

\begin{equation}\label{B_potential}
    V^{AB}_{+}(\boldsymbol{r})+V^{BC}_{-}(\boldsymbol{r})=2V\sum_{j=1}^3\left(\cos{\left(\boldsymbol{g}_j^{AB}\cdot\boldsymbol{r}+\psi\right)}+\cos{\left(\boldsymbol{g}_j^{BC}\cdot\boldsymbol{r}-\psi\right)}\right)
\end{equation}

\begin{figure}[h!]
\centering
\includegraphics[width=0.7\textwidth]{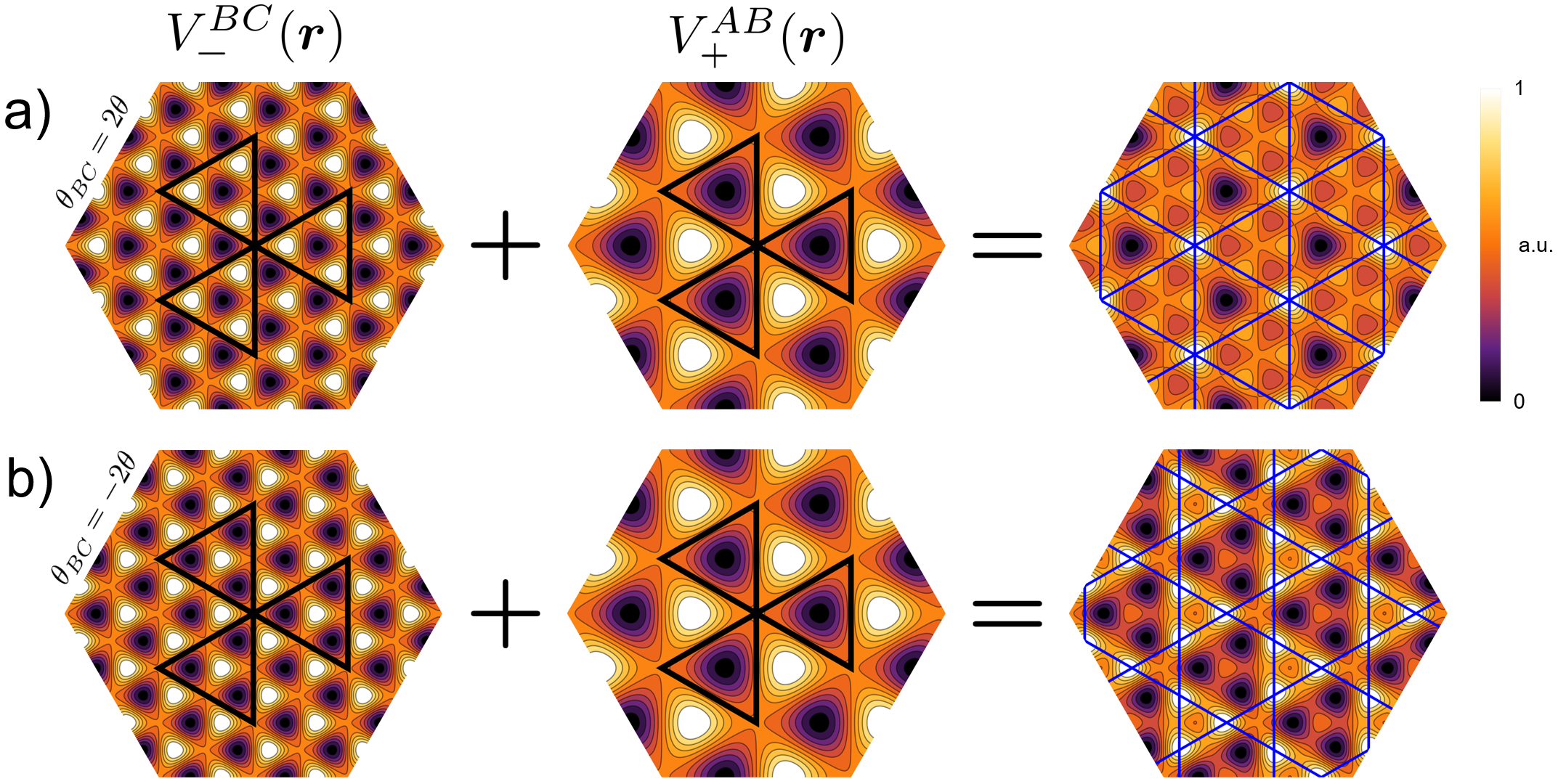}
\caption{\label{realspace_potential_sum_MoTe2} Sum of moir\'e potentials felt by the middle layer of trilayer MoTe$_2$ for a helical twist (a) and an alternating twist (b). Color scales are independently normalized for each plot to emphasize spatial variations. The dark black lines enclose specific triangular regions. These triangular regions of the AB potential lower the potential of points in the BC potential, eliminating the maxima there and leaving a triangular (a) or kagome (b) lattice for the remaining maxima. The blue lines show the resulting lattice of the potential maxima.}
\end{figure}

Knowing that to get a kagome lattice the ratio of the twist angles must be $|\theta_{BC}/\theta_{AB}|=2$, one can choose either a helical twist or alternating twist. We can first check if $\theta_{BC}=2\theta_{AB}$ produces a kagome lattice. The moir\'e potential is given by 
\begin{equation}\label{B_potential1}
    V^{AB}_{+}(\boldsymbol{r})+V^{BC}_{-}(\boldsymbol{r})=2V\sum_{j=1}^3\left(\cos{\left(\boldsymbol{g}_j^{AB}\cdot\boldsymbol{r}+\psi\right)}+\cos{\left(2\boldsymbol{g}_j^{AB}\cdot\boldsymbol{r}-\psi\right)}\right),
\end{equation}
using $\boldsymbol{g}_j^{BC}=2\boldsymbol{g}_j^{AB}$. It turns out that this scheme does not produce a kagome lattice for any value of $\psi$. The lattice with a larger lattice constant always lowers the potential of three of the triangular sublattices, instead of just one, as shown in Fig. \ref{realspace_potential_sum_MoTe2}(a). Consequently, it only produces a triangular lattice for any $\psi$.

However, with an alternating twist, the situation is reversed. If $\theta_{BC}=-2\theta_{AB}$, the potential on layer B is then given by Eq. (\ref{B_potential2}), since $\boldsymbol{g}_j^{BC}=-2\boldsymbol{g}_j^{AB}$. A plot of these two moir\'e potentials adding up to make a kagome potential is in Fig. \ref{realspace_potential_sum_MoTe2}(b). We can also see the kagome potential is breathing as the kagome triangles pointing right have a different potential shape than the triangles pointing left. This is expected as our system has a lower symmetry than the ideal kagome lattice. 

\begin{equation}\label{B_potential2}
    V^{AB}_{+}(\boldsymbol{r})+V^{BC}_{-}(\boldsymbol{r})=2V\sum_{j=1}^3\left(\cos{\left(\boldsymbol{g}_j^{AB}\cdot\boldsymbol{r}+\psi\right)}+\cos{\left(2\boldsymbol{g}_j^{AB}\cdot\boldsymbol{r}+\psi\right)}\right)
\end{equation}

\section{More band structure results}\label{app:more band}
In this section, we present the band structures of our kagome systems with various parameters.





\subsection{Varying twist angles}
\begin{figure}[h!]
\centering
\includegraphics[width=0.6\textwidth]{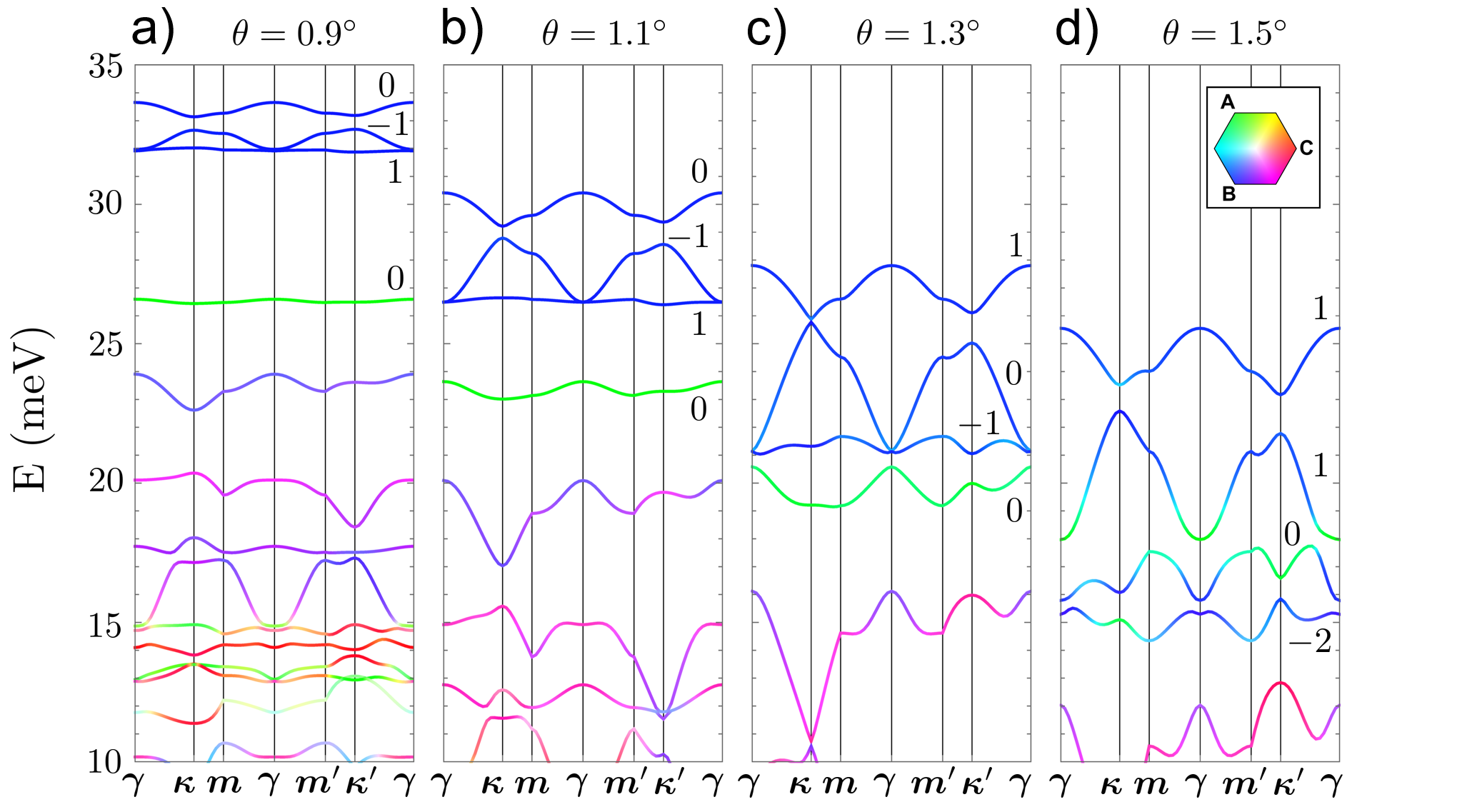}
\caption{\label{FigAngles} Band structure of MoTe$_2$ for different twist angles. The color of the bands indicates the layer polarization, as shown in the legend. The Chern numbers for the top few bands are included next to the bands.}
\end{figure}

We plot the moir\'e band structure of MoTe$_2$ using the parameters in Ref. \cite{WuMacDonald2019} for various twist angles in Fig. \ref{FigAngles}. We can observe that there are discernible kagome bands up to a twist angle of $\theta=1.5^{\circ}$.

\subsection{WSe$_2$ at different angles}
We repeat the same plots of the band structure as a function of $\theta$ using the DFT parameters for WSe$_2$ from \cite{Devakul2021}. 

\begin{figure}[h!]
\centering
\includegraphics[width=0.6\textwidth]{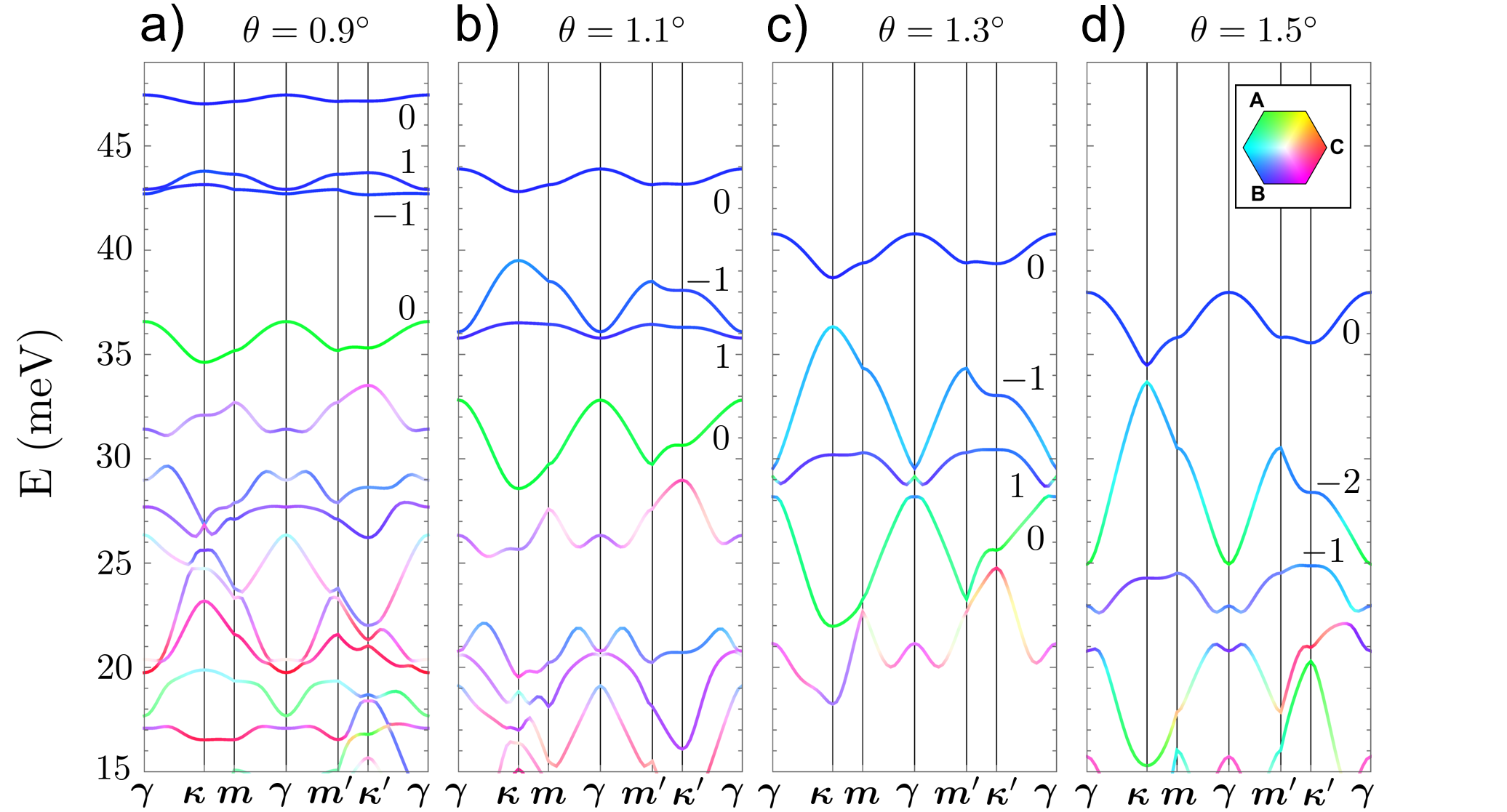}
\caption{\label{FigWSe2} Band structure of WSe$_2$ for different twist angles. The color of the bands indicates the layer polarization, as shown in the legend. The Chern numbers for the top few bands are included next to the bands.}
\end{figure}

\subsection{Different continuum model parameters}

\begin{figure}[h!]
\centering
\includegraphics[width=0.6\textwidth]{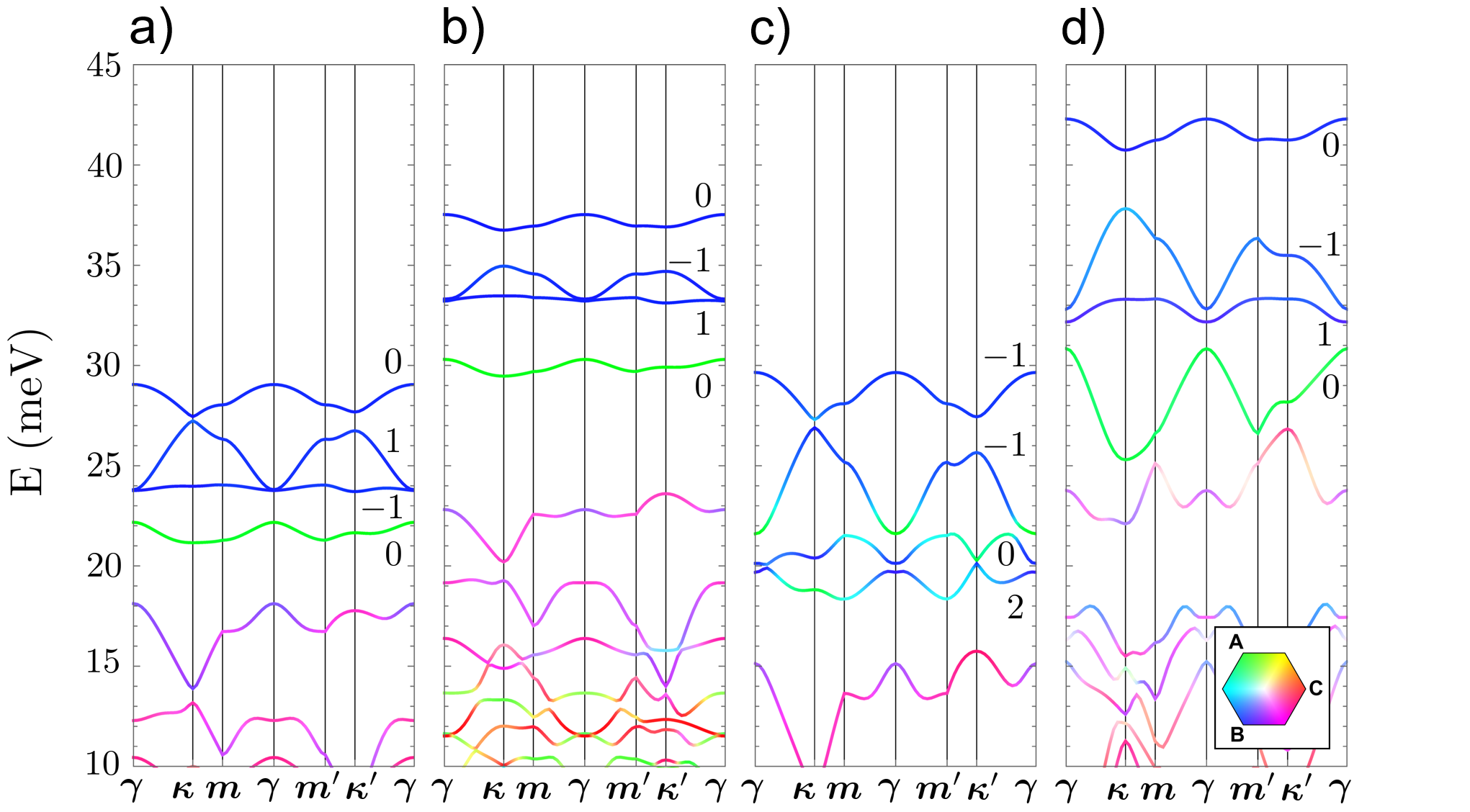}
\caption{\label{Figdiffparams} Band structure with various model parameters. These plots are made with twist angle $\theta=1.2^{\circ}$, and no displacement field. (a) MoTe$_2$ with parameters in Ref. \cite{WuMacDonald2019}. (b) MoTe$_2$ with parameters in Ref. \cite{Xu2024}. (c)  WSe$_2$ with parameters in Ref. \cite{WuMacDonald2019}. (d)  WSe$_2$ with parameters in Ref. \cite{Devakul2021}. We see all of them show reasonable kagome band structures.}
\end{figure}

Here we calculate the moir\'e band structure for our twisting scheme with different parameters from Ref. \cite{WuMacDonald2019, Xu2024, Devakul2021}, shown in Fig. \ref{Figdiffparams}. They show qualitatively similar kagome behaviors. The parameters used are in Table \ref{DFTtable}.

\begin{table}[]
    \centering
    \bgroup
\def\arraystretch{1.4}
\begin{tabular}{|c|c|c|c|c|c|c|c|} 
 \hline
  Material & Reference & Band Structure & $a$ (\AA) & $m$ ($m_e$) & $V$ (meV) & $w$ (meV) & $\psi$ ($^{\circ}$)\\ [0.5ex] 
 \hline
 MoTe$_2$ & \cite{WuMacDonald2019} & Fig. \ref{Figdiffparams}(a) & 3.472 & 0.62 & 8 & -8.5 & -89.6 \\ 
 \hline
 MoTe$_2$ & \cite{Xu2024} & Fig. \ref{Figdiffparams}(b) & 3.52 & 0.62 & 9.2 & -11.2 & -99 \\
 \hline
 WSe$_2$ & \cite{WuMacDonald2019} & Fig. \ref{Figdiffparams}(c) & 3.317 & 0.43 & 8.9 & 9.7 & 91 \\
 \hline
 WSe$_2$ & \cite{Devakul2021} & Fig. \ref{Figdiffparams}(d) & 3.317 & 0.43 & 9 & 18 & 128 \\
 \hline
\end{tabular}
\egroup
\caption{Parameters for the continuum model Hamiltonians that were used to calculate the band structures in Fig. \ref{Figdiffparams}.}
\label{DFTtable}
\end{table}

\subsection{Varying tunneling strength}
 One observation is that the tunneling between layers A and B plays an interesting role, as it pushes the band structure towards the ideal kagome band structure by closing the gap at the $\boldsymbol{\kappa}$-points. The gaps open again and continue to widen as the tunneling is further increased. In this subsection, we treat the tunneling parameter as a knob and plot the band structure as function of the tunneling strength. For simplicity, we fix the twist angle $\theta=1.2^{\circ}$ and use the MoTe$_2$ parameters in Ref. \cite{WuMacDonald2019}.

\begin{figure}[h!]
\centering
\includegraphics[width=0.8\textwidth]{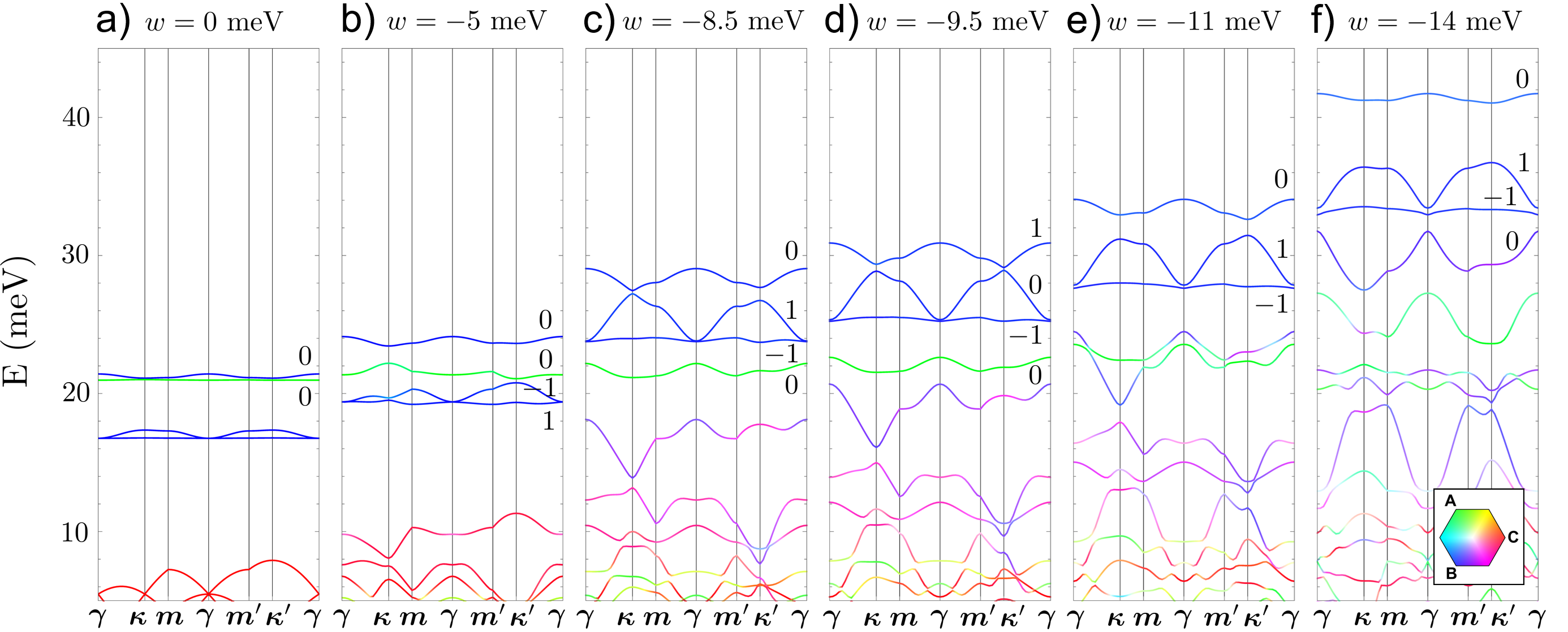}
\caption{\label{FigvaryW} Band structure of MoTe$_2$ (with $\theta=1.2^{\circ}$) for different values of the tunneling parameter $w$. The value of $w$ calculated from DFT is -8.5 meV\cite{WuMacDonald2019}. The color of the bands indicates the layer polarization, as shown in the legend. The Chern numbers for the top few bands are included next to the bands.}
\end{figure}

If we completely turn off the tunneling (result in Fig. \ref{FigvaryW}(a)), we find that the top three blue bands have a large gap at the Dirac point. This signifies a strong breathing lattice. This limit of zero tunneling will correspond to the physical case when we have the three layers starting from AB stacking instead of AA stacking. Because of the strong spin-orbit coupling, valley and spin degree of freedom in TMDs are locked. And if we start from the AB stacking, the tunneling between layers are suppressed due to spin mismatch\cite{Zhang2021}. 

As we dial up the tunneling, we see that the Dirac gaps become smaller (and they are not symmetric for $\boldsymbol{\kappa}$ and $\boldsymbol{\kappa}'$). The Dirac gaps close and then reopen successively as we increase $w$. In the large $w$ regime, again the band structure shows strong breathing (non-ideal) kagome behavior. The band structure in Fig. \ref{FigvaryW} shows that there is a region around $w=-9$ meV where the Dirac gaps are very small. Interestingly, when the tunneling between layers B and C is turned off, the band structure still resembles the kagome band structure, whereas if the tunneling between layers A and B is turned off, the Dirac gaps get much wider and the band structure no longer looks like that of a kagome lattice. Since the breathing kagome band structure is a result of the mismatch of the hoppings between right- and left-pointing triangles, the fact that tunneling to layer A reduces the breathing (for some range of $w$) suggests that tunneling to and from layer A effectively reduces the difference between the hoppings of the right- and left-pointing triangles. 

\section{Wannier orbitals and tight-binding model}\label{app_wannier}

Wannier90\cite{Pizzi2020} was used to find the maximally localized Wannier functions for the top four bands in Fig. \ref{FigvaryW}(c). The parameters for this Hamiltonian are from Ref. \cite{WuMacDonald2019}. First, the Hamiltonian was diagonalized, providing the energies $\varepsilon_{n\boldsymbol{k}}$ and wavefunctions $\ket{\phi_{n\boldsymbol{k}}}$ at each point in the chosen $\boldsymbol{k}$-mesh in the first Brillouin zone. The input to Wannier90 included wavefunction overlaps, initial projections, and the eigenvalues $\varepsilon_{n\boldsymbol{k}}$ for the top four bands.

The wavefunction overlaps are given by Eq. (\ref{Mmn_def}), where the vector $\boldsymbol{b}$ is the vector between adjacent $\boldsymbol{k}$-points in the $\boldsymbol{k}$-mesh.
\begin{equation}\label{Mmn_def}    M_{mn}^{(\boldsymbol{k},\boldsymbol{b})}\equiv\braket{\phi_{m\boldsymbol{k}}|e^{-i\boldsymbol{b}\cdot\hat{\boldsymbol{r}}}|\phi_{n\boldsymbol{k}+\boldsymbol{b}}}
\end{equation}

Wannier90 also needed starting guesses for the Wannier orbitals. These were passed in as projections $A_{mn}(\boldsymbol{k})$, see Eq. (\ref{Amn_def}), of the Bloch states onto the trial localized orbitals $\ket{g_n}$. The trial orbitals chosen are given in Eq. (\ref{gln_def}), where $\boldsymbol{r}_n$ is the initial guess for the $n$th Wannier center, $\sigma_n$ is a guess for the standard deviation of the $n$th Wannier function, and the $f_{ln}$ are layer weights such that $\sum_l|f_{ln}|^2=1$.

\begin{equation}\label{Amn_def}
    A_{mn}(\boldsymbol{k})\equiv\sqrt{N}\braket{\phi_{m\boldsymbol{k}}|g_n}
\end{equation}

\begin{equation}\label{gln_def}
    g_{ln}(\boldsymbol{r})\equiv\braket{\boldsymbol{r},l|g_n}=\frac{f_{ln}}{\sqrt{\pi}\sigma_n}e^{-\frac{1}{2\sigma_n^2}(\boldsymbol{r}-\boldsymbol{r}_n)^2}
\end{equation}

\begin{figure}[h!]
\centering
\includegraphics[width=0.79\textwidth]{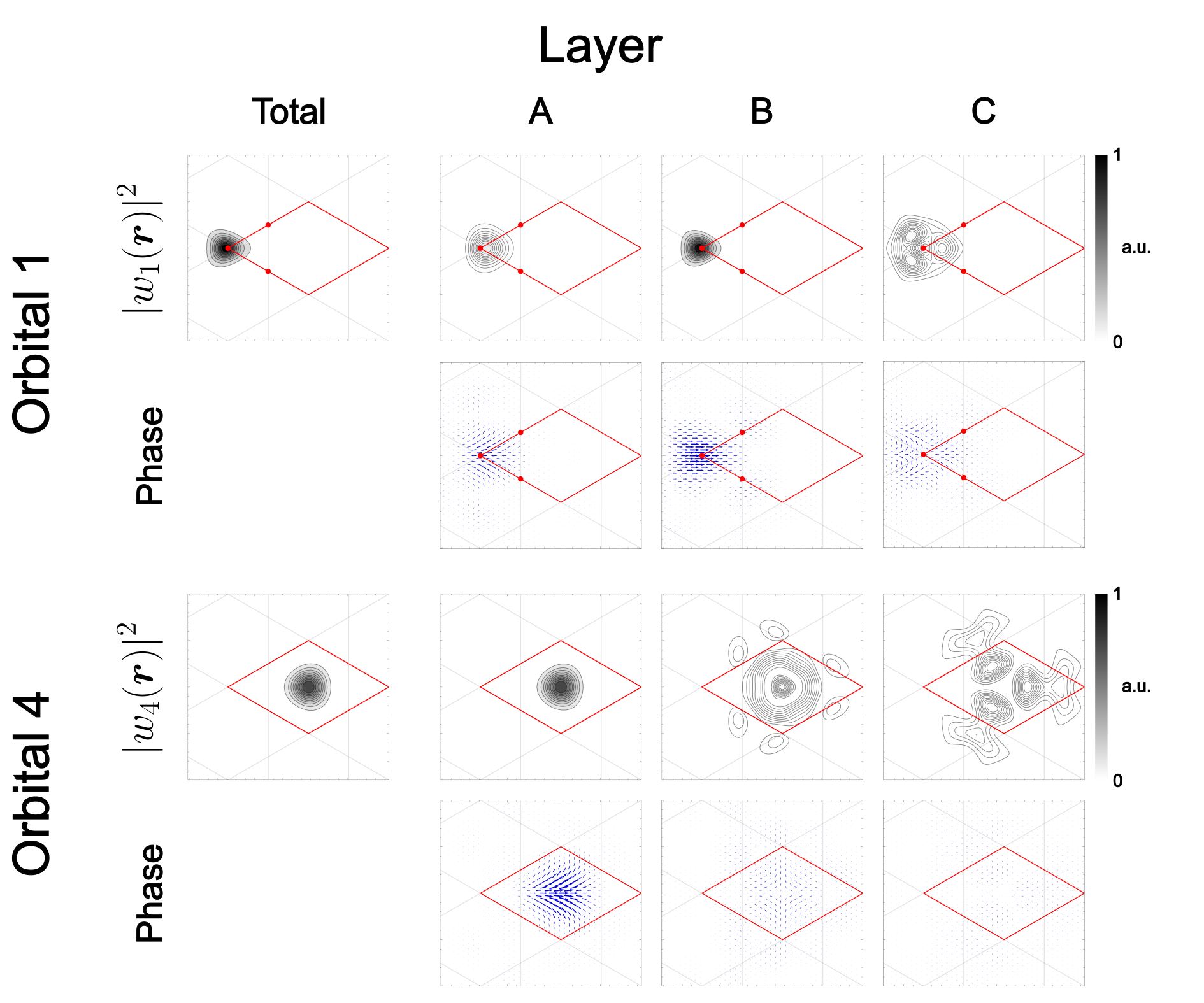}
\caption{\label{FigWannierAll} The total and layer-dependent densities of the maximally localized Wannier functions and the layer-dependent phases. For the phase plots, the arrows point in the direction of the phase, and the length of the vectors is proportional to the square root of the amplitude of the Wannier functions. The Wannier functions for orbitals 2 and 3 can be obtained by rotating the first orbital by $2\pi/3$ and $4\pi/3$ to the other two red sites marked. The red lines outline the moir\'e unit cell.}
\end{figure}

\begin{figure}[h!]
\centering
\includegraphics[width=0.8\textwidth]{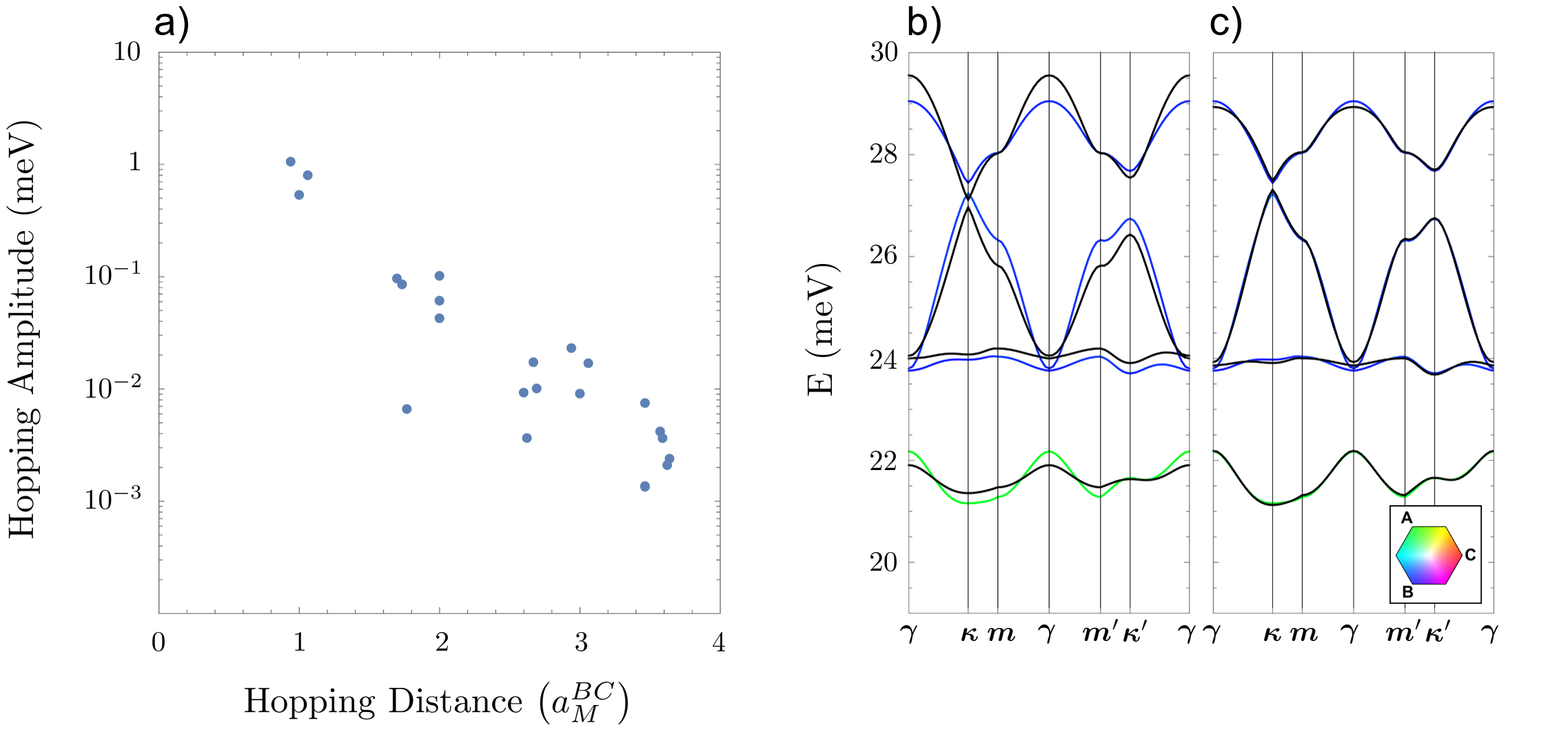}
\caption{\label{FigTB} (a) Hopping amplitudes in the tight-binding model up to 9th nearest neighbors. Here, ``nearest neighbors'' is defined on the ideal kagome lattice since the kagome lattice of the Wannier functions has breathing distortion. We see a good exponential decay of hopping parameters as a function of distance. (b) (c) A comparison of band structures between continuum model calculation and tight-binding model (black) up to first (b) and third (c) nearest neighbors.}
\end{figure}

The maximally localized Wannier functions for the top four bands ($w_i(\boldsymbol{r})$ for $i=1,2,3,4$) are constructed using Wannier90, and the tight-binding model is obtained by extracting the real-space Hamiltonian matrix elements in the Wannier basis. Although the symmetry-adapted Wannier functions option in Wannier90 was not used to enforce $C_3$ symmetry, the resulting Wannier functions and extracted tight-binding parameters exhibited approximate $C_3$ symmetry. The magnitude squared and the phase of the Wannier orbitals are plotted in Fig. \ref{FigWannierAll}. We also present the tight-binding band structure, including hopping terms up to third nearest neighbors, and compare it with the continuum model in Fig. \ref{FigTB}. The tight-binding model reproduces the essential features of the band structure reasonably well. The fluxes acquired from hopping around the symmetry-inequivalent triangular plaquettes are labeled in Fig. \ref{fluxfig}, and the flux values as a function of twist angle are given in Table \ref{fluxtable}.

\begin{table}[]
    \centering
    \begin{tabular}{|c|c|c|c|c|c|c|c|c|}
    \hline
    \rule{0pt}{2.7ex}
    $\theta$ ($^\circ$) & $\phi_K^{\triright}$ ($^\circ$) & $\phi_K^{\trileft}$ ($^\circ$) & $\phi_A^{\triright}$ ($^\circ$) & $\phi_A^{\trileft}$ ($^\circ$) & $|t_{\triright}|$ (meV) & $|t_{\trileft}|$ (meV) & $|t_A|$ (meV) & $\Delta\varepsilon$ (meV) \\[3pt]
    \hline
    0.90 & -13.1 & 8.3 & 41.2 & 80.4 & 0.379 & 0.210 & 0.186 & -6.017 \\
    0.95 & -11.6 & 7.8 & 40.1 & 81.2 & 0.474 & 0.283 & 0.233 & -5.600 \\
    1.00 & -10.0 & 7.2 & 39.0 & 81.9 & 0.578 & 0.368 & 0.285 & -5.210 \\
    1.05 & -8.4 & 6.7 & 38.0 & 82.6 & 0.689 & 0.463 & 0.341 & -4.846 \\
    1.10 & -6.8 & 6.3 & 36.9 & 83.3 & 0.806 & 0.568 & 0.401 & -4.504 \\
    1.15 & -5.2 & 5.9 & 35.8 & 83.9 & 0.928 & 0.681 & 0.466 & -4.181 \\
    1.20 & -3.7 & 5.7 & 34.7 & 84.6 & 1.055 & 0.800 & 0.533 & -3.875 \\
    1.25 & -2.1 & 5.5 & 33.7 & 85.2 & 1.185 & 0.925 & 0.604 & -3.582 \\
    1.30 & -0.5 & 5.4 & 32.6 & 85.8 & 1.320 & 1.054 & 0.677 & -3.299 \\
    1.35 & 1.1 & 5.4 & 31.5 & 86.3 & 1.457 & 1.186 & 0.752 & -3.025 \\
    1.40 & 2.6 & 5.7 & 30.5 & 86.8 & 1.597 & 1.321 & 0.829 & -2.756 \\
    1.45 & 4.2 & 6.1 & 29.4 & 87.1 & 1.739 & 1.455 & 0.906 & -2.489 \\
    1.50 & 5.8 & 6.8 & 28.5 & 87.3 & 1.882 & 1.589 & 0.984 & -2.220 \\
    1.55 & 7.6 & 7.8 & 27.6 & 87.3 & 2.026 & 1.720 & 1.060 & -1.945 \\
    1.60 & 9.4 & 9.3 & 26.9 & 86.8 & 2.170 & 1.846 & 1.134 & -1.659 \\
    \hline
    \end{tabular}
    \caption{\label{fluxtable} Gauge-invariant fluxes through the symmetry-inequivalent triangular plaquettes, magnitude of hopping amplitudes, and difference of onsite energies $\Delta\varepsilon=\varepsilon_{\protect\hexcenterdot} -\varepsilon_{\protect\triright}$ of twisted trilayer MoTe$_2$ as a function of twist angle $\theta$, computed from the Wannier90-derived tight-binding model.}
\end{table}

\begin{table}[]
    \centering
    \begin{tabular}{|c|c|c|c|c|c|c|c|c|c|}
    \hline
    \rule{0pt}{2.7ex}
    $\theta$ ($^\circ$) & $D$ (meV) & $\phi_K^{\triright}$ ($^\circ$) & $\phi_K^{\trileft}$ ($^\circ$) & $\phi_A^{\triright}$ ($^\circ$) & $\phi_A^{\trileft}$ ($^\circ$) & $|t_{\triright}|$ (meV) & $|t_{\trileft}|$ (meV) & $|t_A|$ (meV) & $\Delta\varepsilon$ (meV) \\[3pt]
    \hline
    0.9 & 0 & -13.1 & 8.3 & 41.2 & 80.4 & 0.379 & 0.210 & 0.186 & -6.017 \\
     & 5 & -7.6 & 7.5 & 34.9 & 85.1 & 0.358 & 0.233 & 0.177 & -3.713 \\
     & 10 & -2.8 & 6.8 & 30.3 & 88.4 & 0.336 & 0.263 & 0.179 & -1.439 \\
     & 12 & -1.0 & 6.6 & 28.8 & 89.4 & 0.328 & 0.277 & 0.183 & -0.541 \\
     & 15 & 1.8 & 6.4 & 26.7 & 90.6 & 0.314 & 0.301 & 0.191 & 0.793 \\
     & 18 & 4.7 & 6.2 & 24.8 & 91.6 & 0.301 & 0.330 & 0.203 & 2.109 \\
    \hline
    1.1 & -5 & -14.7 & 5.8 & 44.8 & 78.2 & 0.842 & 0.513 & 0.460 & -6.696 \\
     & -3 & -10.7 & 6.2 & 41.0 & 80.5 & 0.825 & 0.536 & 0.420 & -5.854 \\
     & 0 & -6.8 & 6.3 & 36.9 & 83.3 & 0.806 & 0.568 & 0.401 & -4.504 \\
     & 5 & -1.7 & 6.2 & 31.5 & 87.0 & 0.770 & 0.626 & 0.400 & -2.250 \\
     & 10 & 2.9 & 6.1 & 27.4 & 89.6 & 0.730 & 0.694 & 0.416 & -0.042 \\
     & 15 & 7.7 & 6.2 & 24.0 & 91.4 & 0.685 & 0.776 & 0.446 & 2.104 \\
    \hline
    1.2 & -5 & -10.9 & 5.0 & 41.3 & 80.7 & 1.093 & 0.725 & 0.581 & -6.029 \\
     & -3 & -7.3 & 5.4 & 38.4 & 82.3 & 1.076 & 0.756 & 0.548 & -5.203 \\
     & 0 & -3.7 & 5.7 & 34.7 & 84.6 & 1.055 & 0.800 & 0.533 & -3.875 \\
     & 5 & 1.2 & 5.8 & 29.8 & 87.9 & 1.014 & 0.878 & 0.539 & -1.652 \\
     & 10 & 5.7 & 5.9 & 25.8 & 90.4 & 0.965 & 0.966 & 0.564 & 0.519 \\
     & 14 & 9.5 & 6.1 & 23.1 & 91.7 & 0.920 & 1.046 & 0.596 & 2.204 \\
     & 18 & 13.8 & 6.4 & 20.7 & 92.5 & 0.870 & 1.136 & 0.638 & 3.833 \\
    \hline
    1.3 & -5 & -7.1 & 5.2 & 37.3 & 83.4 & 1.357 & 0.953 & 0.712 & -5.401 \\
     & 0 & -0.5 & 5.4 & 32.6 & 85.8 & 1.320 & 1.054 & 0.677 & -3.299 \\
     & 5 & 4.0 & 5.6 & 28.0 & 88.8 & 1.275 & 1.153 & 0.692 & -1.112 \\
     & 10 & 8.4 & 5.7 & 24.2 & 91.1 & 1.218 & 1.260 & 0.728 & 1.019 \\
     & 15 & 13.2 & 6.1 & 20.9 & 92.6 & 1.149 & 1.379 & 0.780 & 3.069 \\
     & 20 & 19.1 & 6.6 & 18.1 & 93.4 & 1.069 & 1.512 & 0.851 & 5.016 \\
    \hline
    \end{tabular}
    \caption{\label{fluxtableD}Gauge-invariant fluxes through the symmetry-inequivalent triangular plaquettes, magnitude of hopping amplitudes, and difference of onsite energies $\Delta\varepsilon=\varepsilon_{\protect\hexcenterdot} -\varepsilon_{\protect\triright}$ of twisted trilayer MoTe$_2$ as a function of displacement field $D$ for a few different twist angles, computed from the Wannier90-derived tight-binding model.}
\end{table}

\section{Sublattice polarization of the van Hove singularity}\label{app:sublattice}

\begin{figure}[h!]
\centering
\includegraphics[width=0.5\textwidth]{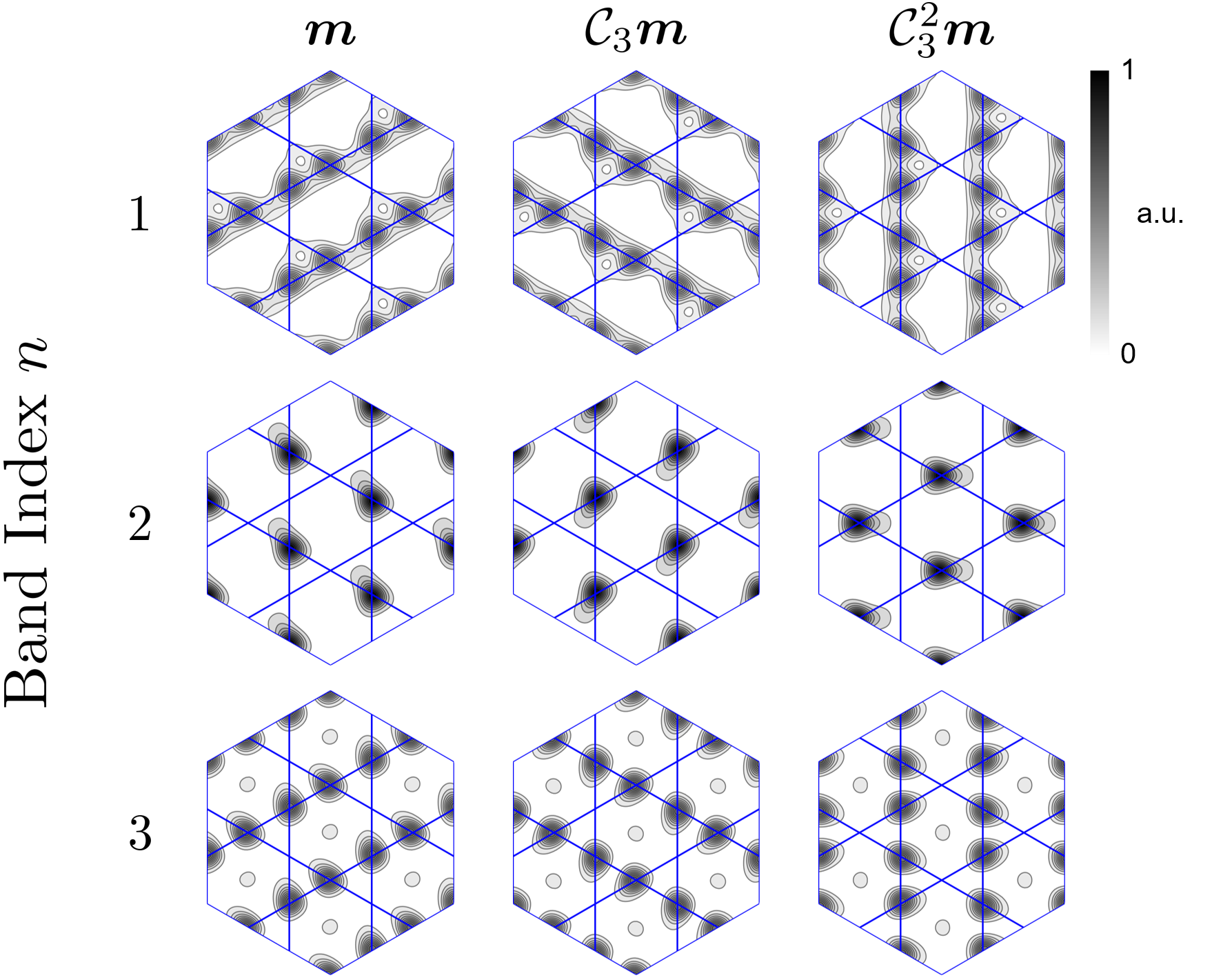}
\caption{\label{realSpaceLayers_MoTe2_mpoints} Real space distribution of $\boldsymbol{m}$-point wavefunctions $|\phi_{n\boldsymbol{k}}(\boldsymbol{r})|^2$ for MoTe$_2$ at $\theta=1.2^{\circ}$. The $\boldsymbol{m}$-points have Van Hove singularities (VHSs). Similar to the ideal kagome lattice, the wavefunctions of the $\boldsymbol{m}$-points have sublattice polarization in real space. Band 2 has a p-type (sublattice pure) VHS, while bands 1 and 3 have m-type (sublattice mixing) VHSs.}
\end{figure}

Plots of the norm squared of the real-space wavefunction $|\phi_{n\boldsymbol{k}}(\boldsymbol{r})|^2=|\braket{\boldsymbol{r}|\phi_{n\boldsymbol{k}}}|^2$ in Fig. \ref{realSpaceLayers_MoTe2_mpoints} show that the wavefunctions for the three inequivalent $\boldsymbol{m}$-points (for the top three bands) are localized on the three sublattices of the kagome lattice. The $\boldsymbol{m}$-points in these bands have Van Hove singularities (VHSs). Similar to the ideal kagome lattice, the wavefunctions of the $\boldsymbol{m}$-points have sublattice polarization in real space. Band 2 has a p-type (sublattice pure) VHS, while bands 1 and 3 have m-type (sublattice mixing) VHSs.

\section{Strong-Coupling Expansion and Effective Spin Hamiltonian}\label{app:strong_coupling}

In this appendix we present the derivation of the effective spin Hamiltonian obtained in the strong-coupling limit of the Hubbard model. The analysis follows standard degenerate perturbation theory and is included for completeness.

\subsection{Degenerate Perturbation Theory}

We consider a Hamiltonian decomposed into a solvable part $H_0\sim U$ and a perturbation $V\sim t$,

\begin{equation*}
    H=H_0+V .
\end{equation*}

Let $P$ denote the projector onto the low-energy manifold and define $Q=I-P$. Inserting $I=P+Q$ and using $PH_0Q\ket{\psi}=0$, we obtain

\begin{equation*}
    EP\ket{\psi}=PH\ket{\psi}=(PHP+PVQ)\ket{\psi}.
\end{equation*}

Projecting instead onto the complementary subspace yields

\begin{align*}
    &EQ\ket{\psi}=QH\ket{\psi}=(QH_0Q+QV)\ket{\psi}\\
    \Rightarrow&(Q(H_0-E)Q)Q\ket{\psi}=-QV\ket{\psi}\\
    \Rightarrow&Q\ket{\psi}=-Q(H_0-E)^{-1}QV\ket{\psi}.
\end{align*}

The final step follows from the fact that $P$ projects onto eigenspaces of $H_0$, implying $[H_0,Q]=[P,H_0]=0$, and hence
$Q(H_0-E)^{-1}Q(Q(H_0-E)Q)=Q$. Substituting this expression for $Q\ket{\psi}$ back into the projected equation for $P\ket{\psi}$ gives

\begin{align*}
    &EP\ket{\psi}=(PHP-PVQ(H_0-E)^{-1}QV)\ket{\psi}\\
    \Rightarrow&EP\ket{\psi}=(PHP-PVQ(H_0-E_0)^{-1}QVP)\ket{\psi}.
\end{align*}

In the final line we retain only terms up to second order in $t/U$. Inserting $I=P+Q$ into the second term generates contributions requiring an additional substitution for $Q\ket{\psi}$, which scale as $O(t^3/U^2)$ and are neglected. Likewise, expanding $E=E_0+\delta E$ produces corrections of order $\delta E\,Q\ket{\psi}\sim O(t^3/U^2)$, since $\delta E\sim O(t^2/U)$ and $Q\ket{\psi}\sim O(t/U)$. These terms are also discarded. The resulting second-order effective Hamiltonian is therefore

\begin{equation*}
    H_{\mathrm{eff}}=E_0P+PVP-PVQ(H_0-E_0)^{-1}QVP .
\end{equation*}

\subsection{Hubbard Model}

We now apply this formalism to the Hubbard model at half filling when $U\gg |t|$ with site-dependent $\varepsilon$ $U$ and bond- and spin-dependent hopping $t$,

\begin{align*}
    &H=H_0+V\\
    &H_0=\sum_i \varepsilon_i n_i+\sum_i U_{i}\,n_{i\uparrow}n_{i\downarrow}\\
    &V=-\sum_{\braket{ij},\sigma}(t_{ij\sigma}c^\dagger_{i\sigma}c_{j\sigma}+\mathrm{h.c.}) .
\end{align*}
    
Here $E_0=\sum_i \varepsilon_i$, and $PVP=0$. Consequently,

\begin{equation*}
    H_{\mathrm{eff}}=E_0 P-PVQ(H_0-E_0)^{-1}QVP=E_0 P-\sum_{\braket{ij}}PV_{ij}Q(H_0-E_0)^{-1}QV_{ij}P .
\end{equation*}

The last sum has the same indices for both $V_{ij}$'s because the $P$'s require that any hopping done in the first $V_{ij}$ is done in reverse in the second $V_{ij}$. Explicitly,

\begin{equation*}
    PV_{ij}Q(H_0-E_0)^{-1}QV_{ij}P=\sum_{\sigma\sigma'}P(t_{ij\sigma}t^*_{ij\sigma'}(U_j+\varepsilon_j-\varepsilon_i)^{-1} c^\dagger_{i\sigma}c_{j\sigma}c^\dagger_{j\sigma'}c_{i\sigma'}+t^*_{ij\sigma}t_{ij\sigma'}(U_i+\varepsilon_i-\varepsilon_j)^{-1} c^\dagger_{j\sigma}c_{i\sigma}c^\dagger_{i\sigma'}c_{j\sigma'})P .
\end{equation*}

These fermionic operators may be rewritten in terms of spin operators defined by

\begin{equation*}
    \vec{S}_i=\frac{1}{2}\sum_{\alpha\beta}c^\dagger_{i\alpha}\vec{\sigma}_{\alpha\beta}c_{i\beta},
\end{equation*}

where $\vec{\sigma}$ denotes the Pauli matrices. The relevant identities are

\begin{align*}
    c^\dagger_{i\sigma}c_{j\sigma}c^\dagger_{j\sigma}c_{i\sigma}&=\tfrac{1}{4}+\sigma\tfrac{1}{2}(S^z_i-S^z_j)-S^z_iS^z_j,\\
    c^\dagger_{i\sigma}c_{j\sigma}c^\dagger_{j\bar{\sigma}}c_{i\bar{\sigma}}&=-S^{\sigma}_iS^{\bar{\sigma}}_j,
\end{align*}

where $\bar{\sigma}$ denotes the opposite spin of $\sigma$. The $\sigma$'s on the right hand side are treated as $+$ or $-$ depending on if $\sigma$ is $\uparrow$ or $\downarrow$ respectively. For example, $S^{\pm}_j=S^x_j\pm i S^y_j$. Substituting these expressions and defining $U_{ij}\equiv2((U_i+\varepsilon_i-\varepsilon_j)^{-1}+(U_j+\varepsilon_j-\varepsilon_i)^{-1})^{-1}$ yields

\begin{align*}
    H_{\mathrm{eff}}=\sum_i \varepsilon_i +\sum_{\braket{ij}}&\left[\frac{2(|t_{ij\uparrow}|^2+|t_{ij\downarrow}|^2)}{U_{ij}}\left(S^z_iS^z_j-\tfrac{1}{4}\right)+\frac{4\mathrm{Re}(t_{ij\uparrow}t^*_{ij\downarrow})}{U_{ij}}\left(S^x_iS^x_j+S^y_iS^y_j\right)\right.\\
    &+\left.\frac{4\mathrm{Im}(t_{ij\uparrow}t^*_{ij\downarrow})}{U_{ij}}\left(\vec{S}_i\times \vec{S}_j\right)_z+\frac{1}{2}(|t_{ij\uparrow}|^2-|t_{ij\downarrow}|^2)((U_i+\varepsilon_i-\varepsilon_j)^{-1}-(U_j+\varepsilon_j-\varepsilon_i)^{-1})(S^z_i-S^z_j)\right].
\end{align*}
The projectors $P$ are redundant and have been dropped without affecting matrix elements. This is because $H_{\mathrm{eff}}$ is constructed to act only within the low-energy manifold, and the resulting spin operators do not connect to states with double occupancy. Imposing time-reversal symmetry, $t_{ij\uparrow}=t^*_{ij\downarrow}=|t_{ij\uparrow}|e^{i\theta_{ij\uparrow}}$, this reduces to

\begin{equation*}
    H_{\mathrm{eff}}=\sum_i \varepsilon_i +\sum_{\langle ij\rangle}\frac{4|t_{ij\uparrow}|^2}{U_{ij}}\left[S_i^zS_j^z-\tfrac{1}{4}+\cos(2\theta_{ij\uparrow})\left(S_i^xS_j^x+S_i^yS_j^y\right)\nonumber+\sin(2\theta_{ij\uparrow})(\vec S_i\times\vec S_j)_z
    \right].
\end{equation*}

When lattice symmetries enforce identical hopping amplitudes on symmetry-related bonds, the effective Hamiltonian can be written in the compact form
\begin{equation}\label{strong_coupling_Heff}
    H_{\mathrm{eff}}=\sum_i \varepsilon_i +\sum_{T}\frac{4|t_{T}|^2}{U_{T}}
    \sum_{\langle ij\rangle\in T}
    \left[
        S_i^z S_j^z- \tfrac{1}{4}
        + \cos(2\theta_{ij})\left(S_i^x S_j^x + S_i^y S_j^y\right)
        + \sin(2\theta_{ij})(\vec{S}_i \times \vec{S}_j)_z
    \right],
\end{equation}
where $T$ labels the symmetry-inequivalent bond classes of the lattice. 
The hopping amplitudes $t_{ij\uparrow}$ are written as $|t_T|e^{i\theta_{ij}}$ for bonds $\langle ij\rangle \in T$. Symmetry requires all bonds in the class $T$ to have the same magnitude $|t_T|$, although their phases $\theta_{ij}$ may differ. The effective interaction strengths $U_T$ are determined by the onsite energies and Hubbard interactions of the two sites connected by each bond.



\twocolumngrid

\bibliographystyle{apsrev4-1} %
\bibliography{MK.bib}

@Preamble{"\providecommand{\noopsort}[1]{}" #
"\providecommand{\singleletter}[1]{#1}%"}

@article{GuertlerMonien2013,
  title   = {Unveiling the Physics of the Doped Phase of the
             \(t\)-\(J\) Model on the Kagome Lattice},
  author  = {Guertler, Siegfried and Monien, Hartmut},
  journal = {Physical Review Letters},
  volume  = {111},
  number  = {9},
  pages   = {097204},
  year    = {2013},
  doi     = {10.1103/PhysRevLett.111.097204}
}

@article{JiangDevereauxKivelson2017,
  title   = {Holon Wigner Crystal in a Lightly Doped Kagome
             Quantum Spin Liquid},
  author  = {Jiang, Hong-Chen and Devereaux, Thomas P. and
             Kivelson, Steven A.},
  journal = {Physical Review Letters},
  volume  = {119},
  number  = {6},
  pages   = {067002},
  year    = {2017},
  doi     = {10.1103/PhysRevLett.119.067002}
}

@article{KoLeeWen2009,
  title   = {Doped Kagome System as Exotic Superconductor},
  author  = {Ko, Wing-Ho and Lee, Patrick A. and Wen, Xiao-Gang},
  journal = {Physical Review B},
  volume  = {79},
  number  = {21},
  pages   = {214502},
  year    = {2009},
  doi     = {10.1103/PhysRevB.79.214502}
}

@article{JiangYaoYang2021,
  title   = {Possible Superconductivity with a Bogoliubov Fermi Surface in a Lightly Doped Kagome \(U(1)\) Spin Liquid},
  author  = {Jiang, Yi-Fan and Yao, Hong and Yang, Fan},
  journal = {Physical Review Letters},
  volume  = {127},
  number  = {18},
  pages   = {187003},
  year    = {2021},
  doi     = {10.1103/PhysRevLett.127.187003}
}

@article{ArracheaCapriottiSorella2004,
  author  = {Arrachea, L. and Capriotti, L. and Sorella, S.},
  title   = {From the triangular to the kagome lattice: 
             Following the footprints of the ordered state},
  journal = {Phys. Rev. B},
  volume  = {69},
  pages   = {224414},
  year    = {2004}
}

@article{Redpath2012,
  author  = {Redpath, T. E. and Hopkinson, J. M. and Leibel, A. A. 
             and Kee, H.-Y.},
  title   = {Dimensional crossover of a frustrated distorted 
             kagome {H}eisenberg model: Application to {FeCrAs}},
  journal = {Phys. Rev. B},
  volume  = {86},
  pages   = {014410},
  year    = {2012}
}

@article{Schaffer2017,
  author = {Schaffer, R. and Huh, Y. and Hwang, K. and Kim, Y. B.},
  title  = {Quantum spin liquid in a breathing kagome lattice},
  journal = {Phys. Rev. B},
  volume = {95},
  pages  = {054410},
  year   = {2017}
}

@article{IqbalPoilblancThomaleBecca2018,
  author = {Iqbal, Y. and Poilblanc, D. and Thomale, R. and Becca, F.},
  title  = {Persistence of the gapless spin liquid in the breathing 
            kagome {H}eisenberg antiferromagnet},
  journal = {Phys. Rev. B},
  volume = {97},
  pages  = {115127},
  year   = {2018}
}

@article{RepellinPollmannHe2017,
  author = {Repellin, C. and He, Y.-C. and Pollmann, F.},
  title  = {Stability of the spin-1/2 kagome ground state with 
            breathing anisotropy},
  journal = {Phys. Rev. B},
  volume = {96},
  pages  = {205124},
  year   = {2017}
}

@article{IqbalPoilblancSchuch2020,
  author = {Iqbal, M. and Poilblanc, D. and Schuch, N.},
  title  = {Gapped {$\mathbb{Z}_{2}$} spin liquid in the breathing 
            kagome {H}eisenberg antiferromagnet},
  journal = {Phys. Rev. B},
  volume = {101},
  pages  = {155141},
  year   = {2020}
}

@article{JahromiOrusPoilblancMila2020,
  author = {Jahromi, S. S. and Or{\'u}s, R. and Poilblanc, D. and Mila, F.},
  title  = {Spin-1/2 kagome {H}eisenberg antiferromagnet with strong 
            breathing anisotropy},
  journal = {SciPost Phys.},
  volume = {9},
  pages  = {092},
  year   = {2020}
}

@article{Messio2010,
  author  = {Messio, L. and Bernu, B. and Lhuillier, C.},
  title   = {Kagome antiferromagnet: A chiral topological spin
             liquid?},
  journal = {Phys. Rev. Lett.},
  volume  = {108},
  pages   = {207204},
  year    = {2012}
}

@article{Reddy2023k,
  title = {Artificial Atoms, Wigner Molecules, and an Emergent Kagome Lattice in Semiconductor Moir\'e Superlattices},
  author = {Reddy, Aidan P. and Devakul, Trithep and Fu, Liang},
  journal = {Phys. Rev. Lett.},
  volume = {131},
  issue = {24},
  pages = {246501},
  numpages = {6},
  year = {2023},
  month = {Dec},
  publisher = {American Physical Society},
  doi = {10.1103/PhysRevLett.131.246501},
  url = {https://link.aps.org/doi/10.1103/PhysRevLett.131.246501}
}

@article{Syozi1951,
    author = {Syôzi, Itiro},
    title = {Statistics of Kagomé Lattice},
    journal = {Progress of Theoretical Physics},
    volume = {6},
    number = {3},
    pages = {306-308},
    year = {1951},
    month = {06},
    issn = {0033-068X},
    doi = {10.1143/ptp/6.3.306},
    url = {https://doi.org/10.1143/ptp/6.3.306},
    eprint = {https://academic.oup.com/ptp/article-pdf/6/3/306/5239621/6-3-306.pdf},
}

@article{Sachdev1992,
  author    = {Subir Sachdev},
  title     = {Kagomé and triangular-lattice Heisenberg antiferromagnets: Ordering from quantum fluctuations and quantum-disordered ground states},
  journal   = {Phys. Rev. B},
  volume    = {45},
  pages     = {12377--12396},
  year      = {1992},
  doi       = {10.1103/PhysRevB.45.12377}
}

@article{Normanreview2016,
  title = {Colloquium: Herbertsmithite and the search for the quantum spin liquid},
  author = {Norman, M. R.},
  journal = {Rev. Mod. Phys.},
  volume = {88},
  issue = {4},
  pages = {041002},
  numpages = {14},
  year = {2016},
  month = {Dec},
  publisher = {American Physical Society},
  doi = {10.1103/RevModPhys.88.041002},
  url = {https://link.aps.org/doi/10.1103/RevModPhys.88.041002}
}

@article{Balents2010,
  author    = {Leon Balents},
  title     = {Spin liquids in frustrated magnets},
  journal   = {Nature},
  volume    = {464},
  pages     = {199--208},
  year      = {2010},
  doi       = {10.1038/nature08917}
}

@article{
Broholm2020,
author = {C. Broholm  and R. J. Cava  and S. A. Kivelson  and D. G. Nocera  and M. R. Norman  and T. Senthil },
title = {Quantum spin liquids},
journal = {Science},
volume = {367},
number = {6475},
pages = {eaay0668},
year = {2020},
doi = {10.1126/science.aay0668},
URL = {https://www.science.org/doi/abs/10.1126/science.aay0668},
eprint = {https://www.science.org/doi/pdf/10.1126/science.aay0668}}

@article{Yan2011,
  author    = {Simeng Yan and David A. Huse and Steven R. White},
  title     = {Spin-liquid ground state of the $S=1/2$ Kagome Heisenberg antiferromagnet},
  journal   = {Science},
  volume    = {332},
  pages     = {1173--1176},
  year      = {2011},
  doi       = {10.1126/science.1201080}
}

@article{Guo2009,
  author    = {Huaiming Guo and Marcel Franz},
  title     = {Topological insulator on the kagome lattice},
  journal   = {Phys. Rev. B},
  volume    = {80},
  pages     = {113102},
  year      = {2009},
  doi       = {10.1103/PhysRevB.80.113102}
}

@article{Tang2011,
  title = {High-Temperature Fractional Quantum Hall States},
  author = {Tang, Evelyn and Mei, Jia-Wei and Wen, Xiao-Gang},
  journal = {Phys. Rev. Lett.},
  volume = {106},
  issue = {23},
  pages = {236802},
  numpages = {4},
  year = {2011},
  month = {Jun},
  publisher = {American Physical Society},
  doi = {10.1103/PhysRevLett.106.236802},
  url = {https://link.aps.org/doi/10.1103/PhysRevLett.106.236802}
}

@article{Xu2015,
  title = {Intrinsic Quantum Anomalous Hall Effect in the Kagome Lattice ${\mathrm{Cs}}_{2}{\mathrm{LiMn}}_{3}{\mathrm{F}}_{12}$},
  author = {Xu, Gang and Lian, Biao and Zhang, Shou-Cheng},
  journal = {Phys. Rev. Lett.},
  volume = {115},
  issue = {18},
  pages = {186802},
  numpages = {5},
  year = {2015},
  month = {Oct},
  publisher = {American Physical Society},
  doi = {10.1103/PhysRevLett.115.186802},
  url = {https://link.aps.org/doi/10.1103/PhysRevLett.115.186802}
}

@article{Kiesel2013,
  author    = {Maximilian L. Kiesel and Christian Platt and Ronny Thomale},
  title     = {Unconventional Fermi surface instabilities in the Kagome Hubbard model},
  journal   = {Phys. Rev. Lett.},
  volume    = {110},
  pages     = {126405},
  year      = {2013},
  doi       = {10.1103/PhysRevLett.110.126405}
}

@article{Ko2009,
  title = {Doped kagome system as exotic superconductor},
  author = {Ko, Wing-Ho and Lee, Patrick A. and Wen, Xiao-Gang},
  journal = {Phys. Rev. B},
  volume = {79},
  issue = {21},
  pages = {214502},
  numpages = {13},
  year = {2009},
  month = {Jun},
  publisher = {American Physical Society},
  doi = {10.1103/PhysRevB.79.214502},
  url = {https://link.aps.org/doi/10.1103/PhysRevB.79.214502}
}

@article{Wang2013,
  title = {Competing electronic orders on kagome lattices at van Hove filling},
  author = {Wang, Wan-Sheng and Li, Zheng-Zhao and Xiang, Yuan-Yuan and Wang, Qiang-Hua},
  journal = {Phys. Rev. B},
  volume = {87},
  issue = {11},
  pages = {115135},
  numpages = {8},
  year = {2013},
  month = {Mar},
  publisher = {American Physical Society},
  doi = {10.1103/PhysRevB.87.115135},
  url = {https://link.aps.org/doi/10.1103/PhysRevB.87.115135}
}

@article{Shores2005,
  author    = {Matthew P. Shores and Emily A. Nytko and Bart M. Bartlett and Daniel G. Nocera},
  title     = {A structurally perfect $S = 1/2$ kagom{\'e} antiferromagnet},
  journal   = {J. Am. Chem. Soc.},
  volume    = {127},
  number    = {39},
  pages     = {13462--13463},
  year      = {2005},
  doi       = {10.1021/ja053891p}
}

@article{Helton2007,
  author    = {J. S. Helton and K. Matan and M. P. Shores and E. A. Nytko and B. M. Bartlett and Y. Yoshida and Y. Takano and A. Suslov and Y. Qiu and J. H. Chung and D. G. Nocera and Y. S. Lee},
  title     = {Spin dynamics of the spin-$\frac{1}{2}$ kagome lattice antiferromagnet ZnCu$_{3}$(OH)$_{6}$Cl$_{2}$},
  journal   = {Phys. Rev. Lett.},
  volume    = {98},
  pages     = {107204},
  year      = {2007},
  doi       = {10.1103/PhysRevLett.98.107204}
}

@article{Mendels2007,
  author    = {Philippe Mendels and Fr{\'e}d{\'e}ric Bert and Marie A. de Vries and Alejandro Olariu and Arnaud Harrison and Florence Duc and Jacques C. Trombe and J. S. Lord and Alain Amato and Claudia Baines},
  title     = {Quantum magnetism in the paratacamite family: towards an ideal kagom{\'e} lattice},
  journal   = {Phys. Rev. Lett.},
  volume    = {98},
  pages     = {077204},
  year      = {2007},
  doi       = {10.1103/PhysRevLett.98.077204}
}

@article{Janson2008,
  author    = {Oleg Janson and Johannes Richter and Helmut Rosner},
  title     = {Modified kagome physics in the natural spin-$\frac{1}{2}$ kagome lattice systems: Kapellasite and haydeeite},
  journal   = {Phys. Rev. Lett.},
  volume    = {101},
  pages     = {106403},
  year      = {2008},
  doi       = {10.1103/PhysRevLett.101.106403}
}

@article{Han2012,
  author    = {Han, Tao-Hsin and Helton, Joshua S. and Chu, Shaoyan and Nocera, Daniel G. and Rodriguez-Rivera, Juan A. and Broholm, Collin and Lee, Young S.},
  title     = {Fractionalized excitations in the spin-liquid state of a kagome-lattice antiferromagnet},
  journal   = {Nature},
  volume    = {492},
  pages     = {406--410},
  year      = {2012},
  doi       = {10.1038/nature11687}
}

@article{Imai2008,
  author    = {Imai, T. and Fu, M. and Han, T. and Nocera, D. G. and Chou, F. C. and Lee, Y. S.},
  title     = {Local Spin Dynamics of the $S=1/2$ Kagome Lattice in ZnCu$_3$(OH)$_6$Cl$_2$},
  journal   = {Physical Review Letters},
  volume    = {100},
  pages     = {077203},
  year      = {2008},
  doi       = {10.1103/PhysRevLett.100.077203}
}

@article{Okamoto2007,
  author    = {Okamoto, Y. and Nohara, M. and Aruga-Katori, H. and Takagi, H.},
  title     = {Spin Liquid State in the $S=1/2$ Kagom\'e Antiferromagnet ZnCu$_3$(OH)$_6$Cl$_2$},
  journal   = {Journal of the Physical Society of Japan},
  volume    = {76},
  number    = {2},
  pages     = {023701},
  year      = {2007},
  doi       = {10.1143/JPSJ.76.023701}
}

@article{Ortiz2019,
  author    = {Brenden R. Ortiz and L{\'i}dia C. Gomes and Jennifer R. Morey and Michal Winiarski and Mitchell Bordelon and John S. Mangum and Iain W. H. Oswald and Jose A. Rodriguez-Rivera and James R. Neilson and Stephen D. Wilson and Elif Ertekin and Tyrel M. McQueen and Eric S. Toberer},
  title     = {New kagome prototype materials: discovery of KV$_3$Sb$_5$, RbV$_3$Sb$_5$, and CsV$_3$Sb$_5$},
  journal   = {Phys. Rev. Materials},
  volume    = {3},
  pages     = {094407},
  year      = {2019},
  doi       = {10.1103/PhysRevMaterials.3.094407}
}

@article{Ortiz2020,
  author    = {Brenden R. Ortiz and Samuel M. L. Teicher and Yong Hu and Julia L. Zuo and Paul M. Sarte and Emily C. Schueller and A. M. Milinda Abeykoon and Matthew J. Krogstad and Stephan Rosenkranz and Raymond Osborn and Ram Seshadri and Leon Balents and Junfeng He and Stephen D. Wilson},
  title     = {CsV$_3$Sb$_5$: A $\mathbb{Z}_2$ topological kagome metal with a superconducting ground state},
  journal   = {Phys. Rev. Lett.},
  volume    = {125},
  pages     = {247002},
  year      = {2020},
  doi       = {10.1103/PhysRevLett.125.247002}
}

@article{Ortiz2021,
  title = {Superconductivity in the ${\mathbb{Z}}_{2}$ kagome metal ${\mathrm{KV}}_{3}{\mathrm{Sb}}_{5}$},
  author = {Ortiz, Brenden R. and Sarte, Paul M. and Kenney, Eric M. and Graf, Michael J. and Teicher, Samuel M. L. and Seshadri, Ram and Wilson, Stephen D.},
  journal = {Phys. Rev. Mater.},
  volume = {5},
  issue = {3},
  pages = {034801},
  numpages = {7},
  year = {2021},
  month = {Mar},
  publisher = {American Physical Society},
  doi = {10.1103/PhysRevMaterials.5.034801},
  url = {https://link.aps.org/doi/10.1103/PhysRevMaterials.5.034801}
}

@Article{Yin2021,
title = {Superconductivity and Normal-State Properties of Kagome Metal RbV$_{3}$Sb$_{5}$ Single Crystals},
journal = {Chin. Phys. Lett.},
volume = {38},
number = {3},
pages = {037403-037403},
year = {2021},
issn = {},
doi = {10.1088/0256-307X/38/3/037403},	
url = {http://cpl.iphy.ac.cn/en/article/doi/10.1088/0256-307X/38/3/037403},
author = {Qiangwei Yin and Zhijun Tu and Chunsheng Gong and Yang Fu  and Shaohua Yan  and Hechang Lei}
}

@article{Jiang2021,
	author = {Jiang, Yu-Xiao and Yin, Jia-Xin and Denner, M. Michael and Shumiya, Nana and Ortiz, Brenden R. and Xu, Gang and Guguchia, Zurab and He, Junyi and Hossain, Md Shafayat and Liu, Xiaoxiong and Ruff, Jacob and Kautzsch, Linus and Zhang, Songtian S. and Chang, Guoqing and Belopolski, Ilya and Zhang, Qi and Cochran, Tyler A. and Multer, Daniel and Litskevich, Maksim and Cheng, Zi-Jia and Yang, Xian P. and Wang, Ziqiang and Thomale, Ronny and Neupert, Titus and Wilson, Stephen D. and Hasan, M. Zahid},
	date = {2021/10/01},
	doi = {10.1038/s41563-021-01034-y},
	id = {Jiang2021},
	isbn = {1476-4660},
	journal = {Nature Materials},
	number = {10},
	pages = {1353--1357},
	title = {Unconventional chiral charge order in kagome superconductor KV3Sb5},
	url = {https://doi.org/10.1038/s41563-021-01034-y},
	volume = {20},
	year = {2021}}

@article{Li2021,
  title = {Observation of Unconventional Charge Density Wave without Acoustic Phonon Anomaly in Kagome Superconductors ${A\mathrm{V}}_{3}{\mathrm{Sb}}_{5}$ ($A=\mathrm{Rb}$, Cs)},
  author = {Li, Haoxiang and Zhang, T. T. and Yilmaz, T. and Pai, Y. Y. and Marvinney, C. E. and Said, A. and Yin, Q. W. and Gong, C. S. and Tu, Z. J. and Vescovo, E. and Nelson, C. S. and Moore, R. G. and Murakami, S. and Lei, H. C. and Lee, H. N. and Lawrie, B. J. and Miao, H.},
  journal = {Phys. Rev. X},
  volume = {11},
  issue = {3},
  pages = {031050},
  numpages = {9},
  year = {2021},
  month = {Sep},
  publisher = {American Physical Society},
  doi = {10.1103/PhysRevX.11.031050},
  url = {https://link.aps.org/doi/10.1103/PhysRevX.11.031050}
}

@article{Zhao2021,
	author = {Zhao, He and Li, Hong and Ortiz, Brenden R. and Teicher, Samuel M. L. and Park, Takamori and Ye, Mengxing and Wang, Ziqiang and Balents, Leon and Wilson, Stephen D. and Zeljkovic, Ilija},
	date = {2021/11/01},
	doi = {10.1038/s41586-021-03946-w},
	id = {Zhao2021},
	isbn = {1476-4687},
	journal = {Nature},
	number = {7884},
	pages = {216--221},
	title = {Cascade of correlated electron states in the kagome superconductor CsV3Sb5},
	url = {https://doi.org/10.1038/s41586-021-03946-w},
	volume = {599},
	year = {2021}}

@article{Chen2021,
	author = {Chen, Hui and Yang, Haitao and Hu, Bin and Zhao, Zhen and Yuan, Jie and Xing, Yuqing and Qian, Guojian and Huang, Zihao and Li, Geng and Ye, Yuhan and Ma, Sheng and Ni, Shunli and Zhang, Hua and Yin, Qiangwei and Gong, Chunsheng and Tu, Zhijun and Lei, Hechang and Tan, Hengxin and Zhou, Sen and Shen, Chengmin and Dong, Xiaoli and Yan, Binghai and Wang, Ziqiang and Gao, Hong-Jun},
	date = {2021/11/01},
	doi = {10.1038/s41586-021-03983-5},
	id = {Chen2021},
	isbn = {1476-4687},
	journal = {Nature},
	number = {7884},
	pages = {222--228},
	title = {Roton pair density wave in a strong-coupling kagome superconductor},
	url = {https://doi.org/10.1038/s41586-021-03983-5},
	volume = {599},
	year = {2021}}

@article{Ye2018,
  author    = {Linda Ye and Mingu Kang and Junwei Liu and Felix von Cube and Christina R. Wicker and Takehito Suzuki and Chris Jozwiak and Aaron Bostwick and Eli Rotenberg and David C. Bell and Liang Fu and Riccardo Comin and Joseph G. Checkelsky},
  title     = {Massive Dirac fermions in a ferromagnetic kagome metal},
  journal   = {Nature},
  volume    = {555},
  pages     = {638--642},
  year      = {2018},
  doi       = {10.1038/nature25987}
}

@article{Kang2020,
  author    = {Mingu Kang and Linda Ye and Shiang Fang and Jhih-Shih You and Abe Levitan and Minyong Han and Jorge I. Facio and Chris Jozwiak and Aaron Bostwick and Eli Rotenberg and Mun K. Chan and Ross D. McDonald and David Graf and Konstantine Kaznatcheev and Elio Vescovo and David C. Bell and Efthimios Kaxiras and Jeroen van den Brink and Manuel Richter and Madhav P. Ghimire and Joseph G. Checkelsky and Riccardo Comin},
  title     = {Dirac fermions and flat bands in the ideal kagome metal FeSn},
  journal   = {Nature Materials},
  volume    = {19},
  pages     = {163--169},
  year      = {2020},
  doi       = {10.1038/s41563-019-0531-0}
}

@article{Li2020FeSn,
  author    = {Li, G. and Wang, H. and Zhang, Q. and et al.},
  title     = {Flat bands and Dirac fermions in the kagome metal FeSn revealed by ARPES},
  journal   = {Advanced Materials},
  volume    = {32},
  pages     = {2001234},
  year      = {2020},
  doi       = {10.1002/adma.202001234}
}

@article{Kuroda2017,
  author    = {Kuroda, K. and Tomiyoshi, S. and Suzuki, M.-T. and Shiomi, Y. and Ota, T. and Miura, Y. and Koizumi, K. and Ueda, K. and Nakatsuji, S.},
  title     = {Evidence for magnetic Weyl fermions in a correlated metal},
  journal   = {Nature Materials},
  volume    = {16},
  pages     = {1090--1095},
  year      = {2017},
  doi       = {10.1038/nmat4925}
}

@article{Nakatsuji2015,
  author    = {Satoru Nakatsuji and Naoki Kiyohara and Tomoya Higo},
  title     = {Large anomalous Hall effect in a non-collinear antiferromagnet at room temperature},
  journal   = {Nature},
  volume    = {527},
  pages     = {212--215},
  year      = {2015},
  doi       = {10.1038/nature15723}
}

@article{Nayak2016,
  author    = {Ajaya K. Nayak and Julia E. Fischer and Yan Sun and Binghai Yan and Julie Karel and Alexander C. Komarek and Chandra Shekhar and Nitesh Kumar and Walter Schnelle and J{\"u}rgen K{\"u}bler and Claudia Felser and Stuart S. P. Parkin},
  title     = {Large anomalous Hall effect driven by a nonvanishing Berry curvature in the noncollinear antiferromagnet Mn$_{3}$Ge},
  journal   = {Science Advances},
  volume    = {2},
  number    = {4},
  pages     = {e1501870},
  year      = {2016},
  doi       = {10.1126/sciadv.1501870}
}

@article{Ikhlas2017,
	author = {Ikhlas, Muhammad and Tomita, Takahiro and Koretsune, Takashi and Suzuki, Michi-To and Nishio-Hamane, Daisuke and Arita, Ryotaro and Otani, Yoshichika and Nakatsuji, Satoru},
	date = {2017/11/01},
	doi = {10.1038/nphys4181},
	id = {Ikhlas2017},
	isbn = {1745-2481},
	journal = {Nature Physics},
	number = {11},
	pages = {1085--1090},
	title = {Large anomalous Nernst effect at room temperature in a chiral antiferromagnet},
	url = {https://doi.org/10.1038/nphys4181},
	volume = {13},
	year = {2017}}

@article{Andrei2021,
title = "The marvels of moir{\'e} materials",
author = "Andrei, {Eva Y.} and Efetov, {Dmitri K.} and Pablo Jarillo-Herrero and MacDonald, {Allan H.} and Mak, {Kin Fai} and T. Senthil and Emanuel Tutuc and Ali Yazdani and Young, {Andrea F.}",
year = "2021",
month = mar,
doi = "10.1038/s41578-021-00284-1",
language = "English (US)",
volume = "6",
pages = "201--206",
journal = "Nature Reviews Materials",
issn = "2058-8437",
publisher = "Nature Publishing Group",
number = "3",
}

@article{Moire2024,
	date = {2024/07/01},
	doi = {10.1038/s41578-024-00698-7},
	id = {ref1},
	isbn = {2058-8437},
	journal = {Nature Reviews Materials},
	number = {7},
	pages = {451--451},
	title = {Moir{\'e}materials keep on giving},
	url = {https://doi.org/10.1038/s41578-024-00698-7},
	volume = {9},
	year = {2024}}

@article{Bistritzer2011,
  author    = {Rafi Bistritzer and Allan H. MacDonald},
  title     = {Moir{\'e} bands in twisted double-layer graphene},
  journal   = {Proc. Natl. Acad. Sci. U.S.A.},
  volume    = {108},
  number    = {30},
  pages     = {12233--12237},
  year      = {2011},
  doi       = {10.1073/pnas.1108174108}
}

@article{Cao2018a,
  author    = {Yuan Cao and Valla Fatemi and Ahmet Demir and Shiang Fang and Spencer L. Tomarken and Jason Y. Luo and Javier D. Sanchez-Yamagishi and Kenji Watanabe and Takashi Taniguchi and Efthimios Kaxiras and Raymond C. Ashoori and Pablo Jarillo-Herrero},
  title     = {Correlated insulator behaviour at half-filling in magic-angle graphene superlattices},
  journal   = {Nature},
  volume    = {556},
  pages     = {80--84},
  year      = {2018},
  doi       = {10.1038/nature26154}
}

@article{Cao2018b,
  author    = {Yuan Cao and Valla Fatemi and Shiang Fang and Kenji Watanabe and Takashi Taniguchi and Efthimios Kaxiras and Pablo Jarillo-Herrero},
  title     = {Unconventional superconductivity in magic-angle graphene superlattices},
  journal   = {Nature},
  volume    = {556},
  pages     = {43--50},
  year      = {2018},
  doi       = {10.1038/nature26160}
}

@article{Serlin2020,
  author    = {Michael Serlin and Christos L. Tschirhart and Houchen Polshyn and Yi Zhang and Jason Zhu and Kenji Watanabe and Takashi Taniguchi and Leon Balents and Andrea F. Young},
  title     = {Intrinsic quantized anomalous Hall effect in a moir{\'e} heterostructure},
  journal   = {Science},
  volume    = {367},
  pages     = {900--903},
  year      = {2020},
  doi       = {10.1126/science.aay5533}
}

@article{Tang2020,
  author    = {Yanhao Tang and Lizhong Li and Tingxin Li and Yang Xu and Song Liu and Katayun Barmak and Kenji Watanabe and Takashi Taniguchi and Allan H. MacDonald and Jie Shan and Kin Fai Mak},
  title     = {Simulation of Hubbard model physics in WSe$_2$/WS$_2$ moir{\'e} superlattices},
  journal   = {Nature},
  volume    = {579},
  pages     = {353--358},
  year      = {2020},
  doi       = {10.1038/s41586-020-2085-3}
}

@article{Regan2020,
  author    = {Emma C. Regan and Danqing Wang and Chenhao Jin and M. Iqbal Bakti Utama and Beini Gao and Xin Wei and Sihan Zhao and Wenyu Zhao and Kentaro Yumigeta and Mark Blei and Johan D. Carlstr{\"o}m and Kenji Watanabe and Takashi Taniguchi and Sefaattin Tongay and Michael F. Crommie and Alex Zettl and Feng Wang},
  title     = {Mott and generalized Wigner crystal states in WSe$_2$/WS$_2$ moir{\'e} superlattices},
  journal   = {Nature},
  volume    = {579},
  pages     = {359--363},
  year      = {2020},
  doi       = {10.1038/s41586-020-2092-4}
}

@article{Chen2020CI,
	author = {Chen, Guorui and Sharpe, Aaron L. and Fox, Eli J. and Zhang, Ya-Hui and Wang, Shaoxin and Jiang, Lili and Lyu, Bosai and Li, Hongyuan and Watanabe, Kenji and Taniguchi, Takashi and Shi, Zhiwen and Senthil, T. and Goldhaber-Gordon, David and Zhang, Yuanbo and Wang, Feng},
	date = {2020/03/01},
	doi = {10.1038/s41586-020-2049-7},
	id = {Chen2020},
	isbn = {1476-4687},
	journal = {Nature},
	number = {7797},
	pages = {56--61},
	title = {Tunable correlated Chern insulator and ferromagnetism in a moir{\'e}superlattice},
	url = {https://doi.org/10.1038/s41586-020-2049-7},
	volume = {579},
	year = {2020}}

@article{Wang2020CI,
	author = {Wang, Lei and Shih, En-Min and Ghiotto, Augusto and Xian, Lede and Rhodes, Daniel A. and Tan, Cheng and Claassen, Martin and Kennes, Dante M. and Bai, Yusong and Kim, Bumho and Watanabe, Kenji and Taniguchi, Takashi and Zhu, Xiaoyang and Hone, James and Rubio, Angel and Pasupathy, Abhay N. and Dean, Cory R.},
	date = {2020/08/01},
	doi = {10.1038/s41563-020-0708-6},
	id = {Wang2020},
	isbn = {1476-4660},
	journal = {Nature Materials},
	number = {8},
	pages = {861--866},
	title = {Correlated electronic phases in twisted bilayer transition metal dichalcogenides},
	url = {https://doi.org/10.1038/s41563-020-0708-6},
	volume = {19},
	year = {2020}}

@article{Li2021CI,
	author = {Li, Hongyuan and Li, Shaowei and Regan, Emma C. and Wang, Danqing and Zhao, Wenyu and Kahn, Salman and Yumigeta, Kentaro and Blei, Mark and Taniguchi, Takashi and Watanabe, Kenji and Tongay, Sefaattin and Zettl, Alex and Crommie, Michael F. and Wang, Feng},
	date = {2021/09/01},
	doi = {10.1038/s41586-021-03874-9},
	id = {Li2021},
	isbn = {1476-4687},
	journal = {Nature},
	number = {7878},
	pages = {650--654},
	title = {Imaging two-dimensional generalized Wigner crystals},
	url = {https://doi.org/10.1038/s41586-021-03874-9},
	volume = {597},
	year = {2021}}

@article{Xu2020CI,
	author = {Xu, Yang and Liu, Song and Rhodes, Daniel A. and Watanabe, Kenji and Taniguchi, Takashi and Hone, James and Elser, Veit and Mak, Kin Fai and Shan, Jie},
	date = {2020/11/01},
	doi = {10.1038/s41586-020-2868-6},
	id = {Xu2020},
	isbn = {1476-4687},
	journal = {Nature},
	number = {7833},
	pages = {214--218},
	title = {Correlated insulating states at fractional fillings of moir{\'e}superlattices},
	url = {https://doi.org/10.1038/s41586-020-2868-6},
	volume = {587},
	year = {2020}}

@article{Chen2019SC,
	author = {Chen, Guorui and Sharpe, Aaron L. and Gallagher, Patrick and Rosen, Ilan T. and Fox, Eli J. and Jiang, Lili and Lyu, Bosai and Li, Hongyuan and Watanabe, Kenji and Taniguchi, Takashi and Jung, Jeil and Shi, Zhiwen and Goldhaber-Gordon, David and Zhang, Yuanbo and Wang, Feng},
	date = {2019/08/01},
	doi = {10.1038/s41586-019-1393-y},
	id = {Chen2019},
	isbn = {1476-4687},
	journal = {Nature},
	number = {7768},
	pages = {215--219},
	title = {Signatures of tunable superconductivity in a trilayer graphene moir{\'e}superlattice},
	url = {https://doi.org/10.1038/s41586-019-1393-y},
	volume = {572},
	year = {2019}}

@article{Liu2020mag,
	author = {Liu, Xiaomeng and Hao, Zeyu and Khalaf, Eslam and Lee, Jong Yeon and Ronen, Yuval and Yoo, Hyobin and Haei Najafabadi, Danial and Watanabe, Kenji and Taniguchi, Takashi and Vishwanath, Ashvin and Kim, Philip},
	date = {2020/07/01},
	doi = {10.1038/s41586-020-2458-7},
	id = {Liu2020},
	isbn = {1476-4687},
	journal = {Nature},
	number = {7815},
	pages = {221--225},
	title = {Tunable spin-polarized correlated states in twisted double bilayer graphene},
	url = {https://doi.org/10.1038/s41586-020-2458-7},
	volume = {583},
	year = {2020}}

@article{Guo2024,
	author = {Guo, Yinjie and Pack, Jordan and Swann, Joshua and Holtzman, Luke and Cothrine, Matthew and Watanabe, Kenji and Taniguchi, Takashi and Mandrus, David G. and Barmak, Katayun and Hone, James and Millis, Andrew J. and Pasupathy, Abhay and Dean, Cory R.},
	date = {2025/01/01},
	doi = {10.1038/s41586-024-08381-1},
	id = {Guo2025},
	isbn = {1476-4687},
	journal = {Nature},
	number = {8047},
	pages = {839--845},
	title = {Superconductivity in 5.0$\,^{\circ}$twisted bilayer WSe2},
	url = {https://doi.org/10.1038/s41586-024-08381-1},
	volume = {637},
	year = {2025}}

@article{Park2021SC,
	author = {Park, Jeong Min and Cao, Yuan and Watanabe, Kenji and Taniguchi, Takashi and Jarillo-Herrero, Pablo},
	date = {2021/02/01},
	doi = {10.1038/s41586-021-03192-0},
	id = {Park2021},
	isbn = {1476-4687},
	journal = {Nature},
	number = {7845},
	pages = {249--255},
	title = {Tunable strongly coupled superconductivity in magic-angle twisted trilayer graphene},
	url = {https://doi.org/10.1038/s41586-021-03192-0},
	volume = {590},
	year = {2021}}

@article{Oh2021,
	author = {Oh, Myungchul and Nuckolls, Kevin P. and Wong, Dillon and Lee, Ryan L. and Liu, Xiaomeng and Watanabe, Kenji and Taniguchi, Takashi and Yazdani, Ali},
	date = {2021/12/01},
	doi = {10.1038/s41586-021-04121-x},
	id = {Oh2021},
	isbn = {1476-4687},
	journal = {Nature},
	number = {7888},
	pages = {240--245},
	title = {Evidence for unconventional superconductivity in twisted bilayer graphene},
	url = {https://doi.org/10.1038/s41586-021-04121-x},
	volume = {600},
	year = {2021}}

@misc{Xia2024SC,
      title={Unconventional superconductivity in twisted bilayer WSe2}, 
      author={Yiyu Xia and Zhongdong Han and Kenji Watanabe and Takashi Taniguchi and Jie Shan and Kin Fai Mak},
      year={2024},
      eprint={2405.14784},
      archivePrefix={arXiv},
      primaryClass={cond-mat.mes-hall},
      url={https://arxiv.org/abs/2405.14784}, 
}

@misc{Han2025,
      title={Signatures of Chiral Superconductivity in Rhombohedral Graphene}, 
      author={Tonghang Han and Zhengguang Lu and Zach Hadjri and Lihan Shi and Zhenghan Wu and Wei Xu and Yuxuan Yao and Armel A. Cotten and Omid Sharifi Sedeh and Henok Weldeyesus and Jixiang Yang and Junseok Seo and Shenyong Ye and Muyang Zhou and Haoyang Liu and Gang Shi and Zhenqi Hua and Kenji Watanabe and Takashi Taniguchi and Peng Xiong and Dominik M. Zumbühl and Liang Fu and Long Ju},
      year={2025},
      eprint={2408.15233},
      archivePrefix={arXiv},
      primaryClass={cond-mat.mes-hall},
      url={https://arxiv.org/abs/2408.15233}, 
}

@article{Li2021QAH,
	author = {Li, Tingxin and Jiang, Shengwei and Shen, Bowen and Zhang, Yang and Li, Lizhong and Tao, Zui and Devakul, Trithep and Watanabe, Kenji and Taniguchi, Takashi and Fu, Liang and Shan, Jie and Mak, Kin Fai},
	date = {2021/12/01},
	doi = {10.1038/s41586-021-04171-1},
	id = {Li2021},
	isbn = {1476-4687},
	journal = {Nature},
	number = {7890},
	pages = {641--646},
	title = {Quantum anomalous Hall effect from intertwined moir{\'e}bands},
	url = {https://doi.org/10.1038/s41586-021-04171-1},
	volume = {600},
	year = {2021}}

@article{Cai2023,
	author = {Cai, Jiaqi and Anderson, Eric and Wang, Chong and Zhang, Xiaowei and Liu, Xiaoyu and Holtzmann, William and Zhang, Yinong and Fan, Fengren and Taniguchi, Takashi and Watanabe, Kenji and Ran, Ying and Cao, Ting and Fu, Liang and Xiao, Di and Yao, Wang and Xu, Xiaodong},
	date = {2023/10/01},
	doi = {10.1038/s41586-023-06289-w},
	id = {Cai2023},
	isbn = {1476-4687},
	journal = {Nature},
	number = {7981},
	pages = {63--68},
	title = {Signatures of fractional quantum anomalous Hall states in twisted MoTe2},
	url = {https://doi.org/10.1038/s41586-023-06289-w},
	volume = {622},
	year = {2023}}

@article{Park2023,
	author = {Park, Heonjoon and Cai, Jiaqi and Anderson, Eric and Zhang, Yinong and Zhu, Jiayi and Liu, Xiaoyu and Wang, Chong and Holtzmann, William and Hu, Chaowei and Liu, Zhaoyu and Taniguchi, Takashi and Watanabe, Kenji and Chu, Jiun-Haw and Cao, Ting and Fu, Liang and Yao, Wang and Chang, Cui-Zu and Cobden, David and Xiao, Di and Xu, Xiaodong},
	date = {2023/10/01},
	doi = {10.1038/s41586-023-06536-0},
	id = {Park2023},
	isbn = {1476-4687},
	journal = {Nature},
	number = {7981},
	pages = {74--79},
	title = {Observation of fractionally quantized anomalous Hall effect},
	url = {https://doi.org/10.1038/s41586-023-06536-0},
	volume = {622},
	year = {2023}}

@article{Zeng2023,
	author = {Zeng, Yihang and Xia, Zhengchao and Kang, Kaifei and Zhu, Jiacheng and Kn{\"u}ppel, Patrick and Vaswani, Chirag and Watanabe, Kenji and Taniguchi, Takashi and Mak, Kin Fai and Shan, Jie},
	date = {2023/10/01},
	doi = {10.1038/s41586-023-06452-3},
	id = {Zeng2023},
	isbn = {1476-4687},
	journal = {Nature},
	number = {7981},
	pages = {69--73},
	title = {Thermodynamic evidence of fractional Chern insulator in moir{\'e}MoTe2},
	url = {https://doi.org/10.1038/s41586-023-06452-3},
	volume = {622},
	year = {2023}}

@article{Lu2024,
	author = {Lu, Zhengguang and Han, Tonghang and Yao, Yuxuan and Reddy, Aidan P. and Yang, Jixiang and Seo, Junseok and Watanabe, Kenji and Taniguchi, Takashi and Fu, Liang and Ju, Long},
	date = {2024/02/01},
	doi = {10.1038/s41586-023-07010-7},
	id = {Lu2024},
	isbn = {1476-4687},
	journal = {Nature},
	number = {8000},
	pages = {759--764},
	title = {Fractional quantum anomalous Hall effect in multilayer graphene},
	url = {https://doi.org/10.1038/s41586-023-07010-7},
	volume = {626},
	year = {2024}}

@article{Foutty2024,
author = {Benjamin A. Foutty  and Carlos R. Kometter  and Trithep Devakul  and Aidan P. Reddy  and Kenji Watanabe  and Takashi Taniguchi  and Liang Fu  and Benjamin E. Feldman },
title = {Mapping twist-tuned multiband topology in bilayer WSe<sub>2</sub>},
journal = {Science},
volume = {384},
number = {6693},
pages = {343-347},
year = {2024},
doi = {10.1126/science.adi4728},
URL = {https://www.science.org/doi/abs/10.1126/science.adi4728},
eprint = {https://www.science.org/doi/pdf/10.1126/science.adi4728}}

@article{Neal2013,
	author = {Neal, Adam T. and Liu, Han and Gu, Jiangjiang and Ye, Peide D.},
	date = {2013/08/27},
	doi = {10.1021/nn402377g},
	isbn = {1936-0851},
	journal = {ACS Nano},
	journal1 = {ACS Nano},
	journal2 = {ACS Nano},
	month = {08},
	number = {8},
	pages = {7077--7082},
	publisher = {American Chemical Society},
	title = {Magneto-transport in MoS2: Phase Coherence, Spin--Orbit Scattering, and the Hall Factor},
	type = {doi: 10.1021/nn402377g},
	url = {https://doi.org/10.1021/nn402377g},
	volume = {7},
	year = {2013},
	year1 = {2013}}

@article{Qu2024,
  title = {Observation of weak localization in dual-gated bilayer $\mathrm{Mo}{\mathrm{S}}_{2}$},
  author = {Qu, Tingyu and Masseroni, Michele and Taniguchi, Takashi and Watanabe, Kenji and \"Ozyilmaz, Barbaros and Ihn, Thomas and Ensslin, Klaus},
  journal = {Phys. Rev. Res.},
  volume = {6},
  issue = {1},
  pages = {013216},
  numpages = {8},
  year = {2024},
  month = {Feb},
  publisher = {American Physical Society},
  doi = {10.1103/PhysRevResearch.6.013216},
  url = {https://link.aps.org/doi/10.1103/PhysRevResearch.6.013216}
}

@article{Scheer2023,
  title = {Kagome and honeycomb flat bands in moir\'e graphene},
  author = {Scheer, Michael G. and Lian, Biao},
  journal = {Phys. Rev. B},
  volume = {108},
  issue = {24},
  pages = {245136},
  numpages = {30},
  year = {2023},
  month = {Dec},
  publisher = {American Physical Society},
  doi = {10.1103/PhysRevB.108.245136},
  url = {https://link.aps.org/doi/10.1103/PhysRevB.108.245136}
}

@article{Scheer2022,
  title = {Magic angles in twisted bilayer graphene near commensuration: Towards a hypermagic regime},
  author = {Scheer, Michael G. and Gu, Kaiyuan and Lian, Biao},
  journal = {Phys. Rev. B},
  volume = {106},
  issue = {11},
  pages = {115418},
  numpages = {40},
  year = {2022},
  month = {Sep},
  publisher = {American Physical Society},
  doi = {10.1103/PhysRevB.106.115418},
  url = {https://link.aps.org/doi/10.1103/PhysRevB.106.115418}
}

@article{WuMacDonald2019,
  title = {Topological Insulators in Twisted Transition Metal Dichalcogenide Homobilayers},
  author = {Wu, Fengcheng and Lovorn, Timothy and Tutuc, Emanuel and Martin, Ivar and MacDonald, A. H.},
  journal = {Phys. Rev. Lett.},
  volume = {122},
  issue = {8},
  pages = {086402},
  numpages = {5},
  year = {2019},
  month = {Feb},
  publisher = {American Physical Society},
  doi = {10.1103/PhysRevLett.122.086402},
  url = {https://link.aps.org/doi/10.1103/PhysRevLett.122.086402}
}

@article{WuMacDonald2018,
  title = {Hubbard Model Physics in Transition Metal Dichalcogenide Moir\'e Bands},
  author = {Wu, Fengcheng and Lovorn, Timothy and Tutuc, Emanuel and MacDonald, A. H.},
  journal = {Phys. Rev. Lett.},
  volume = {121},
  issue = {2},
  pages = {026402},
  numpages = {5},
  year = {2018},
  month = {Jul},
  publisher = {American Physical Society},
  doi = {10.1103/PhysRevLett.121.026402},
  url = {https://link.aps.org/doi/10.1103/PhysRevLett.121.026402}
}

@article{Pizzi2020,
	doi = {10.1088/1361-648x/ab51ff},
	url = {https://doi.org/10.1088%2F1361-648x%2Fab51ff},
	year = 2020,
	month = {jan},
	publisher = {{IOP} Publishing},
	volume = {32},
	number = {16},
	pages = {165902},
	author = {Giovanni Pizzi and Valerio Vitale and Ryotaro Arita and Stefan Blügel and Frank Freimuth and Guillaume G{\'{e}}ranton and Marco Gibertini and Dominik Gresch and Charles Johnson and Takashi Koretsune and Julen Iba{\~{n}}ez-Azpiroz and Hyungjun Lee and Jae-Mo Lihm and Daniel Marchand and Antimo Marrazzo and Yuriy Mokrousov and Jamal I Mustafa and Yoshiro Nohara and Yusuke Nomura and Lorenzo Paulatto and Samuel Ponc{\'{e}} and Thomas Ponweiser and Junfeng Qiao and Florian Thöle and Stepan S Tsirkin and Ma{\l}gorzata Wierzbowska and Nicola Marzari and David Vanderbilt and Ivo Souza and Arash A Mostofi and Jonathan R Yates},
	title = {Wannier90 as a community code: new features and applications},
	journal = {Journal of Physics: Condensed Matter}
}

@article{Kiesel2012,
  title = {Sublattice interference in the kagome Hubbard model},
  author = {Kiesel, Maximilian L. and Thomale, Ronny},
  journal = {Phys. Rev. B},
  volume = {86},
  issue = {12},
  pages = {121105},
  numpages = {4},
  year = {2012},
  month = {Sep},
  publisher = {American Physical Society},
  doi = {10.1103/PhysRevB.86.121105},
  url = {https://link.aps.org/doi/10.1103/PhysRevB.86.121105}
}

@article{Wu2023,
  title = {Sublattice interference promotes pair density wave order in kagome metals},
  author = {Wu, Yi-Ming and Thomale, Ronny and Raghu, S.},
  journal = {Phys. Rev. B},
  volume = {108},
  issue = {8},
  pages = {L081117},
  numpages = {6},
  year = {2023},
  month = {Aug},
  publisher = {American Physical Society},
  doi = {10.1103/PhysRevB.108.L081117},
  url = {https://link.aps.org/doi/10.1103/PhysRevB.108.L081117}
}

@article{Wang2013VanHove,
  title = {Competing electronic orders on kagome lattices at van Hove filling},
  author = {Wang, Wan-Sheng and Li, Zheng-Zhao and Xiang, Yuan-Yuan and Wang, Qiang-Hua},
  journal = {Phys. Rev. B},
  volume = {87},
  issue = {11},
  pages = {115135},
  numpages = {8},
  year = {2013},
  month = {Mar},
  publisher = {American Physical Society},
  doi = {10.1103/PhysRevB.87.115135},
  url = {https://link.aps.org/doi/10.1103/PhysRevB.87.115135}
}

@article{Jia2024,
  title = {Moir\'e fractional Chern insulators. I. First-principles calculations and continuum models of twisted bilayer ${\mathrm{MoTe}}_{2}$},
  author = {Jia, Yujin and Yu, Jiabin and Liu, Jiaxuan and Herzog-Arbeitman, Jonah and Qi, Ziyue and Pi, Hanqi and Regnault, Nicolas and Weng, Hongming and Bernevig, B. Andrei and Wu, Quansheng},
  journal = {Phys. Rev. B},
  volume = {109},
  issue = {20},
  pages = {205121},
  numpages = {29},
  year = {2024},
  month = {May},
  publisher = {American Physical Society},
  doi = {10.1103/PhysRevB.109.205121},
  url = {https://link.aps.org/doi/10.1103/PhysRevB.109.205121}
}

@article{Mao2024,
	author = {Mao, Ning and Xu, Cheng and Li, Jiangxu and Bao, Ting and Liu, Peitao and Xu, Yong and Felser, Claudia and Fu, Liang and Zhang, Yang},
	da = {2024/08/03},
	doi = {10.1038/s42005-024-01754-y},
	id = {Mao2024},
	isbn = {2399-3650},
	journal = {Communications Physics},
	number = {1},
	pages = {262},
	title = {Transfer learning relaxation, electronic structure and continuum model for twisted bilayer MoTe2},
	ty = {JOUR},
	url = {https://doi.org/10.1038/s42005-024-01754-y},
	volume = {7},
	year = {2024}}

@article{Xia2025,
	author = {Xia, Li-Qiao and de la Barrera, Sergio C. and Uri, Aviram and Sharpe, Aaron and Kwan, Yves H. and Zhu, Ziyan and Watanabe, Kenji and Taniguchi, Takashi and Goldhaber-Gordon, David and Fu, Liang and Devakul, Trithep and Jarillo-Herrero, Pablo},
	da = {2025/02/01},
	doi = {10.1038/s41567-024-02731-6},
	id = {Xia2025},
	isbn = {1745-2481},
	journal = {Nature Physics},
	number = {2},
	pages = {239--244},
	title = {Topological bands and correlated states in helical trilayer graphene},
	ty = {JOUR},
	url = {https://doi.org/10.1038/s41567-024-02731-6},
	volume = {21},
	year = {2025}}

@article{Devakul2023,
author = {Trithep Devakul  and Patrick J. Ledwith  and Li-Qiao Xia  and Aviram Uri  and Sergio C. de la Barrera  and Pablo Jarillo-Herrero  and Liang Fu },
title = {Magic-angle helical trilayer graphene},
journal = {Science Advances},
volume = {9},
number = {36},
pages = {eadi6063},
year = {2023},
doi = {10.1126/sciadv.adi6063},
URL = {https://www.science.org/doi/abs/10.1126/sciadv.adi6063},
eprint = {https://www.science.org/doi/pdf/10.1126/sciadv.adi6063}}

@article{Mak2010,
  title={Atomically Thin MoS2: A New Direct-Gap Semiconductor},
  author={Mak, Kin F. and Lee, Changgu and Hone, James and Shan, Jie and Heinz, Tony F.},
  journal={Physical Review Letters},
  volume={105},
  number={13},
  pages={136805},
  year={2010},
  publisher={American Physical Society}
}

@article{Zhao2013,
  title={Lattice dynamics in mono- and few-layer sheets of WS2 and WSe2},
  author={Zhao, Wei and Ghorannevis, Zand and Amara, Khalil K and Pang, Jian-Rong and Toh, Mei and Zhang, Yong and Kloc, Christoph and Tan, Ping-Heng and Eda, Goki},
  journal={Nanoscale},
  volume={5},
  number={20},
  pages={9677--9683},
  year={2013},
  publisher={Royal Society of Chemistry}
}

@article{Zhang2021,
  title = {SU(4) Chiral Spin Liquid, Exciton Supersolid, and Electric Detection in Moir\'e Bilayers},
  author = {Zhang, Ya-Hui and Sheng, D. N. and Vishwanath, Ashvin},
  journal = {Phys. Rev. Lett.},
  volume = {127},
  issue = {24},
  pages = {247701},
  numpages = {6},
  year = {2021},
  month = {Dec},
  publisher = {American Physical Society},
  doi = {10.1103/PhysRevLett.127.247701},
  url = {https://link.aps.org/doi/10.1103/PhysRevLett.127.247701}
}

@article{Xu2024,
	author = {Cheng Xu and Jiangxu Li and Yong Xu and Zhen Bi and Yang Zhang},
	doi = {10.1073/pnas.2316749121},
	eprint = {https://www.pnas.org/doi/pdf/10.1073/pnas.2316749121},
	journal = {Proceedings of the National Academy of Sciences},
	number = {8},
	pages = {e2316749121},
	title = {Maximally localized Wannier functions, interaction models, and fractional quantum anomalous Hall effect in twisted bilayer MoTe<sub>2</sub>},
	url = {https://www.pnas.org/doi/abs/10.1073/pnas.2316749121},
	volume = {121},
	year = {2024}}

@article{Devakul2021,
author={Devakul, Trithep
and Cr{\'e}pel, Valentin
and Zhang, Yang
and Fu, Liang},
title={Magic in twisted transition metal dichalcogenide bilayers},
journal={Nature Communications},
year={2021},
month={Nov},
day={18},
volume={12},
number={1},
pages={6730},
issn={2041-1723},
doi={10.1038/s41467-021-27042-9},
url={https://doi.org/10.1038/s41467-021-27042-9}
}

@article{Reddy2023,
  title = {Fractional quantum anomalous Hall states in twisted bilayer ${\mathrm{MoTe}}_{2}$ and ${\mathrm{WSe}}_{2}$},
  author = {Reddy, Aidan P. and Alsallom, Faisal and Zhang, Yang and Devakul, Trithep and Fu, Liang},
  journal = {Phys. Rev. B},
  volume = {108},
  issue = {8},
  pages = {085117},
  numpages = {10},
  year = {2023},
  month = {Aug},
  publisher = {American Physical Society},
  doi = {10.1103/PhysRevB.108.085117},
  url = {https://link.aps.org/doi/10.1103/PhysRevB.108.085117}
}

@article{Wang2024,
  title = {Fractional Chern Insulator in Twisted Bilayer ${\mathrm{MoTe}}_{2}$},
  author = {Wang, Chong and Zhang, Xiao-Wei and Liu, Xiaoyu and He, Yuchi and Xu, Xiaodong and Ran, Ying and Cao, Ting and Xiao, Di},
  journal = {Phys. Rev. Lett.},
  volume = {132},
  issue = {3},
  pages = {036501},
  numpages = {6},
  year = {2024},
  month = {Jan},
  publisher = {American Physical Society},
  doi = {10.1103/PhysRevLett.132.036501},
  url = {https://link.aps.org/doi/10.1103/PhysRevLett.132.036501}
}

@article{Morales2023,
  title = {Pressure-enhanced fractional Chern insulators along a magic line in moir\'e transition metal dichalcogenides},
  author = {Morales-Dur\'an, Nicol\'as and Wang, Jie and Schleder, Gabriel R. and Angeli, Mattia and Zhu, Ziyan and Kaxiras, Efthimios and Repellin, C\'ecile and Cano, Jennifer},
  journal = {Phys. Rev. Res.},
  volume = {5},
  issue = {3},
  pages = {L032022},
  numpages = {6},
  year = {2023},
  month = {Aug},
  publisher = {American Physical Society},
  doi = {10.1103/PhysRevResearch.5.L032022},
  url = {https://link.aps.org/doi/10.1103/PhysRevResearch.5.L032022}
}

@article{Nakatsuji2025,
author={Nakatsuji, Naoto and Kawakami, Takuto and Tateishi, Hayato and Kato, Koichiro and Koshino, Mikito},
title={Moir{\'e} band engineering in twisted trilayer WSe2},
journal={Communications Materials},
year={2025},
month={Nov},
day={28},
volume={6},
number={1},
pages={274},
issn={2662-4443},
doi={10.1038/s43246-025-00996-9},
url={https://doi.org/10.1038/s43246-025-00996-9}
}

\end{document}